\RequirePackage{lineno}

\documentclass[aps,pro,twocolumn,superscriptaddress,showpacs]{revtex4-1}

\usepackage{amssymb}
\usepackage{amsmath}
\usepackage{multirow}

\usepackage{graphicx, subfigure}

\begin{document}

\title{Upper bound on neutrino mass based on T2K neutrino timing measurements}





\newcommand{\INSTC}{\affiliation{University of Alberta, Centre for Particle Physics, Department of Physics, Edmonton, Alberta, Canada}}
\newcommand{\INSTEE}{\affiliation{University of Bern, Albert Einstein Center for Fundamental Physics, Laboratory for High Energy Physics (LHEP), Bern, Switzerland}}
\newcommand{\INSTFE}{\affiliation{Boston University, Department of Physics, Boston, Massachusetts, U.S.A.}}
\newcommand{\INSTD}{\affiliation{University of British Columbia, Department of Physics and Astronomy, Vancouver, British Columbia, Canada}}
\newcommand{\INSTGA}{\affiliation{University of California, Irvine, Department of Physics and Astronomy, Irvine, California, U.S.A.}}
\newcommand{\INSTI}{\affiliation{IRFU, CEA Saclay, Gif-sur-Yvette, France}}
\newcommand{\INSTGB}{\affiliation{University of Colorado at Boulder, Department of Physics, Boulder, Colorado, U.S.A.}}
\newcommand{\INSTFG}{\affiliation{Colorado State University, Department of Physics, Fort Collins, Colorado, U.S.A.}}
\newcommand{\INSTFH}{\affiliation{Duke University, Department of Physics, Durham, North Carolina, U.S.A.}}
\newcommand{\INSTBA}{\affiliation{Ecole Polytechnique, IN2P3-CNRS, Laboratoire Leprince-Ringuet, Palaiseau, France }}
\newcommand{\INSTEF}{\affiliation{ETH Zurich, Institute for Particle Physics, Zurich, Switzerland}}
\newcommand{\INSTEG}{\affiliation{University of Geneva, Section de Physique, DPNC, Geneva, Switzerland}}
\newcommand{\INSTDG}{\affiliation{H. Niewodniczanski Institute of Nuclear Physics PAN, Cracow, Poland}}
\newcommand{\INSTCB}{\affiliation{High Energy Accelerator Research Organization (KEK), Tsukuba, Ibaraki, Japan}}
\newcommand{\INSTED}{\affiliation{Institut de Fisica d'Altes Energies (IFAE), Bellaterra (Barcelona), Spain}}
\newcommand{\INSTEC}{\affiliation{IFIC (CSIC \& University of Valencia), Valencia, Spain}}
\newcommand{\INSTEI}{\affiliation{Imperial College London, Department of Physics, London, United Kingdom}}
\newcommand{\INSTGF}{\affiliation{INFN Sezione di Bari and Universit\`a e Politecnico di Bari, Dipartimento Interuniversitario di Fisica, Bari, Italy}}
\newcommand{\INSTBE}{\affiliation{INFN Sezione di Napoli and Universit\`a di Napoli, Dipartimento di Fisica, Napoli, Italy}}
\newcommand{\INSTBF}{\affiliation{INFN Sezione di Padova and Universit\`a di Padova, Dipartimento di Fisica, Padova, Italy}}
\newcommand{\INSTBD}{\affiliation{INFN Sezione di Roma and Universit\`a di Roma ``La Sapienza'', Roma, Italy}}
\newcommand{\INSTEB}{\affiliation{Institute for Nuclear Research of the Russian Academy of Sciences, Moscow, Russia}}
\newcommand{\INSTHA}{\affiliation{Kavli Institute for the Physics and Mathematics of the Universe (WPI), Todai Institutes for Advanced Study, University of Tokyo, Kashiwa, Chiba, Japan}}
\newcommand{\INSTCC}{\affiliation{Kobe University, Kobe, Japan}}
\newcommand{\INSTCD}{\affiliation{Kyoto University, Department of Physics, Kyoto, Japan}}
\newcommand{\INSTEJ}{\affiliation{Lancaster University, Physics Department, Lancaster, United Kingdom}}
\newcommand{\INSTFC}{\affiliation{University of Liverpool, Department of Physics, Liverpool, United Kingdom}}
\newcommand{\INSTFI}{\affiliation{Louisiana State University, Department of Physics and Astronomy, Baton Rouge, Louisiana, U.S.A.}}
\newcommand{\INSTJ}{\affiliation{Universit\'e de Lyon, Universit\'e Claude Bernard Lyon 1, IPN Lyon (IN2P3), Villeurbanne, France}}
\newcommand{\INSTHB}{\affiliation{Michigan State University, Department of Physics and Astronomy,  East Lansing, Michigan, U.S.A.}}
\newcommand{\INSTCE}{\affiliation{Miyagi University of Education, Department of Physics, Sendai, Japan}}
\newcommand{\INSTDF}{\affiliation{National Centre for Nuclear Research, Warsaw, Poland}}
\newcommand{\INSTFJ}{\affiliation{State University of New York at Stony Brook, Department of Physics and Astronomy, Stony Brook, New York, U.S.A.}}
\newcommand{\INSTGJ}{\affiliation{Okayama University, Department of Physics, Okayama, Japan}}
\newcommand{\INSTCF}{\affiliation{Osaka City University, Department of Physics, Osaka, Japan}}
\newcommand{\INSTGG}{\affiliation{Oxford University, Department of Physics, Oxford, United Kingdom}}
\newcommand{\INSTBB}{\affiliation{UPMC, Universit\'e Paris Diderot, CNRS/IN2P3, Laboratoire de Physique Nucl\'eaire et de Hautes Energies (LPNHE), Paris, France}}
\newcommand{\INSTGC}{\affiliation{University of Pittsburgh, Department of Physics and Astronomy, Pittsburgh, Pennsylvania, U.S.A.}}
\newcommand{\INSTFA}{\affiliation{Queen Mary University of London, School of Physics and Astronomy, London, United Kingdom}}
\newcommand{\INSTE}{\affiliation{University of Regina, Department of Physics, Regina, Saskatchewan, Canada}}
\newcommand{\INSTGD}{\affiliation{University of Rochester, Department of Physics and Astronomy, Rochester, New York, U.S.A.}}
\newcommand{\INSTBC}{\affiliation{RWTH Aachen University, III. Physikalisches Institut, Aachen, Germany}}
\newcommand{\INSTFB}{\affiliation{University of Sheffield, Department of Physics and Astronomy, Sheffield, United Kingdom}}
\newcommand{\INSTDI}{\affiliation{University of Silesia, Institute of Physics, Katowice, Poland}}
\newcommand{\INSTEH}{\affiliation{STFC, Rutherford Appleton Laboratory, Harwell Oxford,  and  Daresbury Laboratory, Warrington, United Kingdom}}
\newcommand{\INSTCH}{\affiliation{University of Tokyo, Department of Physics, Tokyo, Japan}}
\newcommand{\INSTBJ}{\affiliation{University of Tokyo, Institute for Cosmic Ray Research, Kamioka Observatory, Kamioka, Japan}}
\newcommand{\INSTCG}{\affiliation{University of Tokyo, Institute for Cosmic Ray Research, Research Center for Cosmic Neutrinos, Kashiwa, Japan}}
\newcommand{\INSTGI}{\affiliation{Tokyo Metropolitan University, Department of Physics, Tokyo, Japan}}
\newcommand{\INSTF}{\affiliation{University of Toronto, Department of Physics, Toronto, Ontario, Canada}}
\newcommand{\INSTB}{\affiliation{TRIUMF, Vancouver, British Columbia, Canada}}
\newcommand{\INSTG}{\affiliation{University of Victoria, Department of Physics and Astronomy, Victoria, British Columbia, Canada}}
\newcommand{\INSTDJ}{\affiliation{University of Warsaw, Faculty of Physics, Warsaw, Poland}}
\newcommand{\INSTDH}{\affiliation{Warsaw University of Technology, Institute of Radioelectronics, Warsaw, Poland}}
\newcommand{\INSTFD}{\affiliation{University of Warwick, Department of Physics, Coventry, United Kingdom}}
\newcommand{\INSTGE}{\affiliation{University of Washington, Department of Physics, Seattle, Washington, U.S.A.}}
\newcommand{\INSTGH}{\affiliation{University of Winnipeg, Department of Physics, Winnipeg, Manitoba, Canada}}
\newcommand{\INSTEA}{\affiliation{Wroclaw University, Faculty of Physics and Astronomy, Wroclaw, Poland}}
\newcommand{\INSTH}{\affiliation{York University, Department of Physics and Astronomy, Toronto, Ontario, Canada}}

\INSTC
\INSTEE
\INSTFE
\INSTD
\INSTGA
\INSTI
\INSTGB
\INSTFG
\INSTFH
\INSTBA
\INSTEF
\INSTEG
\INSTDG
\INSTCB
\INSTED
\INSTEC
\INSTEI
\INSTGF
\INSTBE
\INSTBF
\INSTBD
\INSTEB
\INSTHA
\INSTCC
\INSTCD
\INSTEJ
\INSTFC
\INSTFI
\INSTJ
\INSTHB
\INSTCE
\INSTDF
\INSTFJ
\INSTGJ
\INSTCF
\INSTGG
\INSTBB
\INSTGC
\INSTFA
\INSTE
\INSTGD
\INSTBC
\INSTFB
\INSTDI
\INSTEH
\INSTCH
\INSTBJ
\INSTCG
\INSTGI
\INSTF
\INSTB
\INSTG
\INSTDJ
\INSTDH
\INSTFD
\INSTGE
\INSTGH
\INSTEA
\INSTH

\author{K.\,Abe}\INSTBJ
\author{J.\,Adam}\INSTFJ
\author{H.\,Aihara}\INSTCH\INSTHA
\author{T.\,Akiri}\INSTFH
\author{C.\,Andreopoulos}\INSTEH\INSTFC
\author{S.\,Aoki}\INSTCC
\author{A.\,Ariga}\INSTEE
\author{S.\,Assylbekov}\INSTFG
\author{D.\,Autiero}\INSTJ
\author{M.\,Barbi}\INSTE
% ...

\collaboration{The T2K Collaboration}\noaffiliation

\date{\today}

\begin{abstract}
The Tokai to Kamioka (T2K) long-baseline neutrino experiment consists of a muon neutrino beam, produced at the J-PARC accelerator, a near detector complex and a large 295~km distant far detector. 
The present work utilizes the T2K event timing measurements at the near and far detectors to study neutrino time of flight as function of derived neutrino energy. 
Under the assumption of a relativistic relation between energy and time of flight, constraints on the neutrino rest mass can be derived.
The sub-GeV neutrino beam in conjunction with timing precision of order tens of ns provide sensitivity to neutrino mass in the few MeV/$c^2$ range. 
We study the distribution of relative arrival times of muon and electron neutrino candidate events at the T2K far detector as a function of neutrino energy. 
 The 90\% C.L. upper limit on the mixture of neutrino mass eigenstates represented in the data sample is found to be  
m$_{\nu}^2 < 5.6 \,  {\mathrm MeV^2/}c^4$. 
\end{abstract}

\pacs{14.60.Pq,14.60.Lm,29.40.ka}

\maketitle

\section{INTRODUCTION}

Over the past one and a half decades a variety of experiments unequivocally demonstrated that neutrinos change flavor as they travel from
their source towards a suitably located detector \cite{SK,SNO,K2K,KamLAND,Borexino,MINOS,T2K,DayaBay,DCHOOZ,Reno}. Neutrino oscillations were observed in atmospheric \cite{SK} 
and solar neutrinos \cite{SNO}. The oscillations were confirmed by accelerator \cite{K2K,MINOS,T2K} and reactor based experiments \cite{KamLAND} and have now been studied in a variety 
of different channels \cite{Borexino,DCHOOZ,DayaBay,Reno} with improving accuracy. 

These experimental results imply that neutrinos have non-zero rest mass. 
However, our knowledge of the neutrino masses remains limited and
the determining the values of the neutrino masses remains one of the most important problems of particle physics.
To date only limits on neutrino masses exist.
Neutrino oscillation measurement determine the mass squared differences and thereby provide lower
bounds on the heavier neutrino mass eigenstates.
A variety of experimental approaches with different sensitivities derive upper limits on neutrino mass:

Measurements of pion at rest decay parameters are muon based and find upper limits for the square root of 
$m^{2(eff)}_{\nu_{\mu}} \equiv \sum_i |U_{\mu i}|^2 \, m^2_{\nu_i}$, where $U_{\mu i}$ represent elements of the PMNS neutrino mixing matrix,
 of order 0.2~MeV/$c^2$ (90\% C.L.) \cite{numass_piondecay, numass_piondecay_historical}.
Limits on $m^{(eff)}_{\nu_{\mu}}$ can also be derived from nucleosynthesis in combination with cosmology and are found to be $m^{(eff)}_{\nu_{\mu}} \lesssim 0.2$ MeV/$c^2$
(90 \% C.L.) \cite{numass_nucleosyn,RPP2012}.

Neutrino time of flight (TOF) measurements \cite{nu_TOF_historical} at an accelerator based neutrino long baseline experiments \cite{MINOS_TOF} select $\nu_{\mu}$ and $\overline{\nu}_{\mu}$ candidate events to derive a  99\% C.L. upper limit on $m^{(eff)}_{\nu_{\mu}/ \overline{\nu}_{\mu}} <$  50 MeV/$c^2$.
A neutrino TOF based upper mass limit using $\overline{\nu}_{e}$
 had previously been derived from SN1987A data \cite{numass_SN1987A}.
Upper limits  on the square root of $m^{2(eff)}_{\overline{\nu}_{e}} \equiv \sum_i |\overline{U}_{e i}|^2 \, m^2_{\nu_i}$ 
 are found to be $m^{(eff)}_{\overline{\nu}_{e}} <5.8$~eV/$c^2$ (95\% C.L.).

Tritium beta decay experiments are sensitive to the same quantity $m^{(eff)}_{\nu_{e}} $ by measuring the end-point of the beta spectrum.  
These direct neutrino mass measurement experiments tend to have the best limits without having to rely on assumptions about model parameters.
Limits are found to be $m^{(eff)}_{\nu_{e}}  <2.0$~eV/$c^2$ (95\% C.L.) \cite{numass_3beta}.\\
Neutrino-less double beta decay experiments produce limits for  $m^{(eff)}_{\nu_{e}}$ which are an order of magnitude more stringent but depend on the 
assumption that neutrinos are Majorana particles and rely on very uncertain nuclear matrix element calculations.\\
The sum of all neutrino masses $m_{tot}$ is constrained by measurements of the cosmological background radiation 
to $m_{tot} < 0.2$~eV/$c^2$ (95\% C.L.) \cite{RPP2012} but limits are strongly dependent on assumptions about the cosmological 
parameters.\\
The Tokai to Kamioka (T2K) experiment allows estimation of the effective neutrino mass associated with muon and electron neutrinos using 
their relative TOF (RTOF) between the near and far detectors.
The mass is derived from two quantities: the
energy of the neutrino candidate events and their TOF between near and far detectors relative to the mean TOF for the most energetic neutrino candidate events. 
At lower energies the neutrino rest mass represents a larger fraction of the total neutrino energy which leads to a larger neutrino TOF if the rest mass is sufficiently large. 
The T2K experiment offers a competitive opportunity for such a measurement because of its very short beam bunches and a relatively low mean neutrino beam energy for which the magnitude of the delay due to the neutrino rest mass increases. 
The analysis uses a sample of charged current quasi-elastic (CCQE) neutrino candidate events 
at the far detector for which neutrino energies can be derived with good accuracy. We combine the electron and muon neutrino CCQE samples to maximize statistics.
The near detector is used to measure the neutrino bunch times which represent the start time for the neutrino TOF.\\

Section~\ref{sec:T2KExp} provides an overview of the T2K components and is followed by a detailed description of the hardware setup of timing components in 
section~\ref{sec:T2KTimSys}. In section~\ref{sec:DataSelec} we describe the data selection at the near and far detector and demonstrate good stability of all components of the timing system. Section~\ref{sec:DataAnal} starts with an overview of the data analysis before we demonstrate the analysis performance using toy data sets, applying it to the experimentally recorded data and describing the treatment of systematic uncertainties. Results are given in section~\ref{sec:Results} followed by a summary 
in section~\ref{sec:Summary}.

\section{THE T2K EXPERIMENT}
\label{sec:T2KExp}

The T2K experiment~\cite{t2knim} uses a 30 GeV proton
beam from the J-PARC accelerator facility. The experiment
combines (1) a muon neutrino beam line, 
(2) the near detector complex, which is located 280 m downstream of the neutrino production target and measures the neutrino beam, which constrains the neutrino flux and cross sections, and (3) the far detector, Super-Kamiokande (SK), which detects neutrinos at a distance of $L=295$ km from the target. The neutrino beam axis
is directed 2.5$^\circ$ away from SK producing a narrow-band $\nu_{\mu}$ beam~\cite{PhysRevD.87.012001} at the far detector
with an energy peak at  
$E_\nu \approx 0.6$~GeV.
This corresponds to the first minimum of the $\nu_\mu$ survival probability at SK, thus
enhances the sensitivity to neutrino oscillations
and reduces backgrounds from higher-energy neutrino interactions at SK.

The J-PARC main ring (MR) accelerator produces a fast-extracted  proton beam. 
 The primary beam line has 21 electrostatic beam position monitors, 19 secondary emission monitors and an optical transition radiation monitor to measure the beam profile, and five current transformers (CT) which measure the proton current before a graphite target.
Pions and kaons produced in the target decay in the secondary beam line, 
which contains three focusing horns and a 96-m-long decay tunnel. This is followed by a beam dump and
a set of muon monitors (MUMON)~\cite{mumon}.\\
The near detector complex contains an on-axis Interactive Neutrino Grid detector (INGRID)~\cite{Abe2012} and an off-axis magnetic detector, ND280.
A more detailed detector description is published elsewhere~\cite{Abe:2011ks}.  
The INGRID detector has 14 seven-ton iron-scintillator tracker modules arranged in a 10-m horizontal by 10-m vertical crossed array. This detector provides high-statistics monitoring of the beam intensity, direction, profile, and stability.  The off-axis detector is enclosed in a 0.2-T magnet that contains 
a subdetector optimized to measure $\pi^0$s (P$\O$D)~\cite{Assylbekov201248},
three time projection chambers (TPC1,2,3)~\cite{Abgrall:2010hi} 
alternating with two one-ton fine-grained detectors (FGD1,2)~\cite{Amaudruz:2012pe}, 
and an electromagnetic calorimeter (ECal)~\cite{allan2013electromagnetic} that surrounds the TPC, FGD, and P$\O$D detectors. A side muon range detector (SMRD)~\cite{Aoki:2012mf} consists of 2008 scintillator counters sandwiched between the iron plates which make up the ND280 magnet flux return yokes. 
Each counter is read out by two photosensors, one on each side of the counter. 
The SMRD identifies muons that exit or stop in the magnet steel when the path length exceeds the energy loss range. 
The SK water Cherenkov far detector~\cite{Ashie:2005ik} has a 22.5 kt fiducial volume within a cylindrical inner detector (ID) instrumented with 11129 inward facing 20-inch phototubes. Surrounding the ID is a 2-meter wide outer detector (OD) with 1885 outward-facing 8-inch phototubes. A Global Positioning System with $<$150 ns precision \cite{OT-GPS_paper} synchronizes the timing between SK events and the J-PARC  beam spill. 

These results are based on the data accumulated in four periods: Run I (January-June 2010), Run II (November 2010-March 2011), Run III (January-June 2012) and Run IV (October 2012 - May 2013). 
The proton beam power on the target steadily increased from Run I, reaching 250 kW with about $1.2 \times 10^{14}$ protons per pulse on the target by the end of Run IV.
The total neutrino beam exposure on the SK detector corresponds to an integrated $6.57 \times 10^{20}$ protons on target (POT).

\section{The T2K timing system}
\label{sec:T2KTimSys}

The SK and J-PARC time synchronization systems are almost identical, with only minor differences.
Each system includes two independent GPS receivers from different manufacturers (GPS1 and GPS2), a
Rubidium atomic clock, and a custom Local Time Clock (LTC) board which serves as time keeper and matches
signals to specific times, that is it  generates and distributes timestamps.
The GPS receivers provide time data every second and the
Rubidium atomic clock provides a stable precision time base for the LTC, which generates times 
every 10 ns. The 1 pulse per second (PPS) signals from the GPS receivers are used to reset the fine-scale
counters in the LTC. The time data are integrated in the LTC module, which communicates directly with
the local detector DAQ system.

A beam trigger signal is generated and linked to the MR radio frequency (RF) to ensure synchronization with the proton beam. This trigger signal is provided to the power supplies of both the magnetic horn and the fast extraction (FX) kicker magnet 3 ms before the beam extraction. The beam trigger signal is distributed through an optical fiber from the MR control room to the neutrino beam line control room (NU1) where it's arrival time is measured by the LTC. 

At J-PARC the LTC is located alongside the GPS receivers at NU1 and uses a 100 MHz (10ns) master clock rate. 
The time stamped trigger signal is then distributed to provide the ADC gate timing for the beam line CTs, other proton beam monitors and MUMON. 

The beam timing is monitored by CT1 which is the most upstream of the neutrino beam line proton-beam monitors.   The relative time between the CT1 signal and the edge of CT1's ADC gate is set to an arbitrary value of 1~$\mu$s to account for changes in the beam arrival time after tuning of accelerator parameters.
  The uncertainty of the 1~$\mu$s delay is less than 50~ns. In the present study we measure near detector hit times and far detector event times relative to CT1 beam signals to eliminate any drifts in the proton bunch arrival times. Hence the accuracy is better than 50~ns. 
All electronics delays are fixed during beam operation such that the relative timing between CT1, MUMON and ND280 are constant for a given run.

One copy of the LTC trigger signal is sent via optical fiber to the neutrino monitor building (NM) to provide the beam trigger for ND280 and INGRID. 
The master clock module in the ND280 electronics timestamps ND280 hits relative to the LTC time signal.

The SMRD time stamps are recorded by frontend boards which are controlled by FPGAs. The frontend board timestamps the data with a precision of 2.5 ns. The threshold to generate a timestamp is programmable from 0 to 5 p.e. and is set to 2 p.e. in the case of the SMRD. The ADC and timestamp data are assembled by the FPGA and sent to a backend board for data concentration and buffering. More details are given in \cite{Aoki:2012mf, t2knim}.

The beam spill trigger timestamp produced at NU1 is also sent to the SK DAQ server via a virtual private network (VPN). 
An acceptance window of 1 ms width is used to identify T2K beam related events using a software trigger
filter applied to the buffered event data stream. The 1 ms wide time window is centered on the time of the received beam trigger, offset by the light-speed travel
time for the 295 km distance between J-PARC and SK.

At SK,  GPS receivers are located outside the mine entrance 
and the LTC module is located in the Central Electronics Hut on top of the SK detector.
The receivers and LTC are identical to the ones used at NU1. 
second
At SK, event time stamping is done using a 60kHz (17$\mu$s) master clock frequency.
The SK hardware trigger module (TRG) receives the 60kHz signal and counts the
number of cycles since the previous 1 PPS provided by the GPS. 
The TRG hardware requires a 50 MHz signal which is supplied by a Rubidium clock whose 10 MHz output has been converted.
Multiple PMT hits in a trigger of each 17$\mu$s period share a common GPS time stamp. 
The SK front-end electronics 
records the time of each PMT hit as the difference from the latest 17$\mu$s clock signal
using a charge to time converter (QTC) which has 0.52 ns resolution. 
The QTC counts are reset by the 60kHz clock.

\section{DATA SELECTION AND TIMING STABILITY}
\label{sec:DataSelec} 

At the near site the primary goal is to measure the timing of the neutrino beam bunch structure. For this purpose neutrino events with interaction vertices inside the ND280 and the surrounding soil (Òsand eventsÓ) can be used. The near detector used for this analysis is the SMRD because it provides a high statistics neutrino candidate sample 
due to the large mass of the magnet yokes in which it is embedded.The SMRD allows a straight forward selection of detector hits to maximize beam related hits while keeping
noise hits at a minimum.

At the far detector two beam neutrino candidate data samples are identified, one to extract a neutrino mass limit and a second one to perform system timing stability checks. 
Beam neutrino candidate events at SK are selected by requiring that events are fully contained (FC) inside the detector fiducial volume and within a 1~msec time window centered on the expected arrival time of the beam spill.
 The first sample consists of FC fiducial volume (FCFV) events with only one muon-like or electron-like ring. This data sample corresponds to CCQE neutrino candidate events and represents the data set to extract an upper bound on neutrino mass. 
 The second sample consists of FC events with a visible energy above 2~GeV and does not contain any events from the first sample. It is used to characterize the combined timing uncertainty of the SK detector and the GPS system. The former data sample is used to study energy dependent RTOF effects as it allows an accurate estimate of neutrino energy on an event-by-event basis.

For event selection at both the near and far detector, we check the stability of the time stamping mechanism with respect to proton beam bunches as measured by CT1.
Since neutrino candidate events of interest are related to the proton beam bunches, the event times in the near and far detector 
are coupled to the times when the proton beam bunches arrive on target. Any detector specific timing instabilities would appear as a large fluctuation in the time difference between the detector event and the CT1 signal. 
The residual timing distributions of selected events at SK and the SMRD are used to determine all timing uncertainties, which are relevant for a neutrino RTOF analysis. 
Since the SK events are time stamped with respect to the SK-GPS time and the beam trigger time uses the NU1-GPS system as a time reference, good relative stability between the SK and NU1 GPS time references are required. Hence, the distribution of residual event times with respect to the center of the nearest beam bunch for neutrino candidate events at SK is a measure of the combination of the stability of SK and the relative stability of the NU1 and SK GPS reference times.

\subsection{Near detector data selection and timing}

The selected SMRD data sample is based on good beam spills for which all ND280 data quality cut criteria are satisfied \cite{T2K} regardless of whether there is a SK event or not. The total SMRD data set for run periods I, II, III and IV represents 1.65$\times$10$^{19}$ POT, 7.89$\times$10$^{19}$ POT, 1.57$\times$10$^{20}$ POT and 3.25$\times$10$^{20}$ POT. 
The instantaneous beam intensities during these running periods were such that on average, a few SMRD hits are observed per spill.
SMRD hits are selected if both photosensors which read out a single counter from opposite ends, each create a signal above 4.5 photo-electrons (p.e.) in coincidence.
The mean light yield for a perpendicularly penetrating muon amounts to 40 p.e.
 No additional event reconstruction is applied. 
 
\begin{figure}[t]
\includegraphics[width=\linewidth]{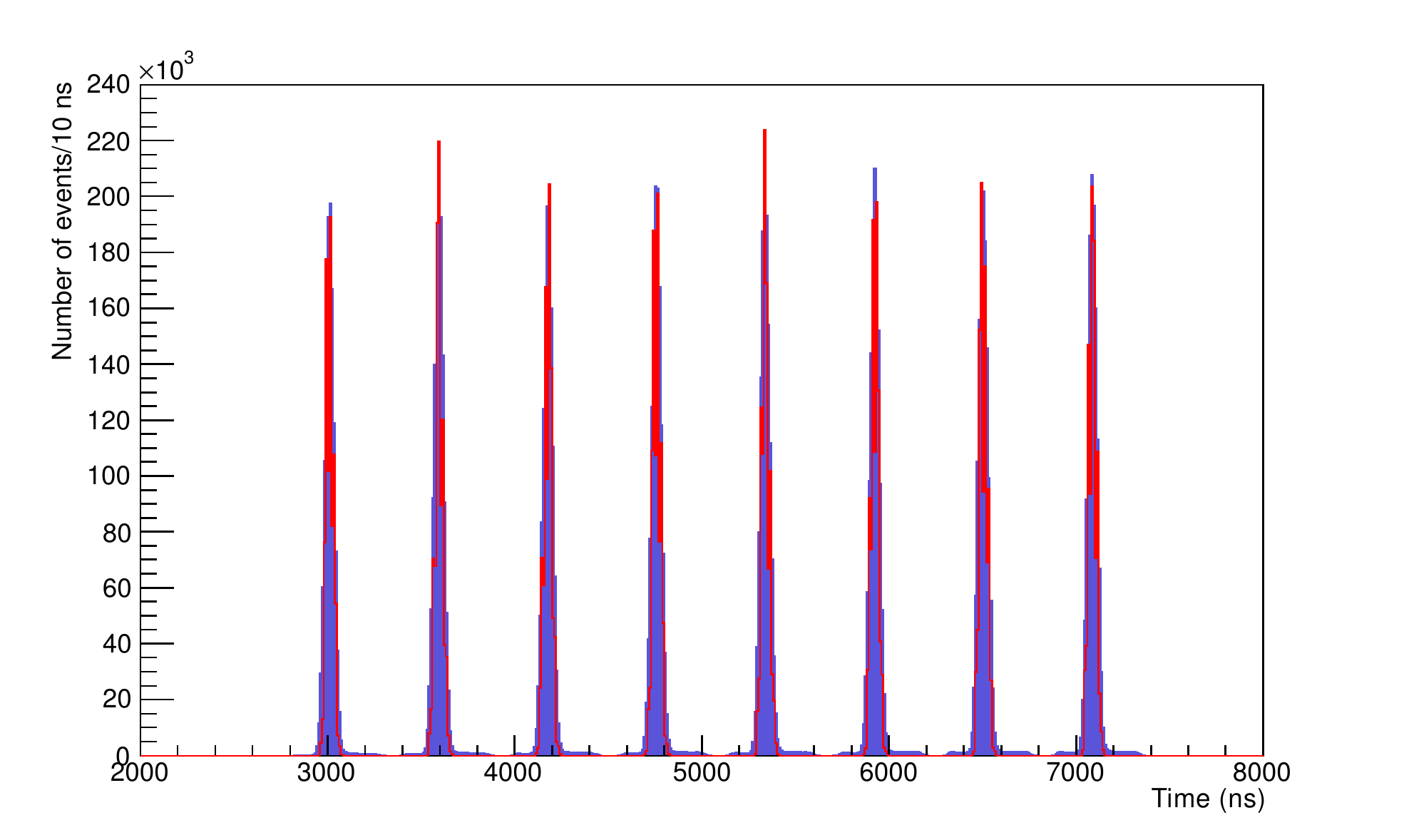}
\caption{\label{fig:8bunch}  Time distribution of the hits selected in SMRD (red: t$_{SMRD}$) and CT1 (black: t$_{CT1}$) integrated over run period III.  The 8 bunch structure of the beam is clearly visible.}
\end{figure}

\begin{figure}[t]
\includegraphics[width=0.5\textwidth]{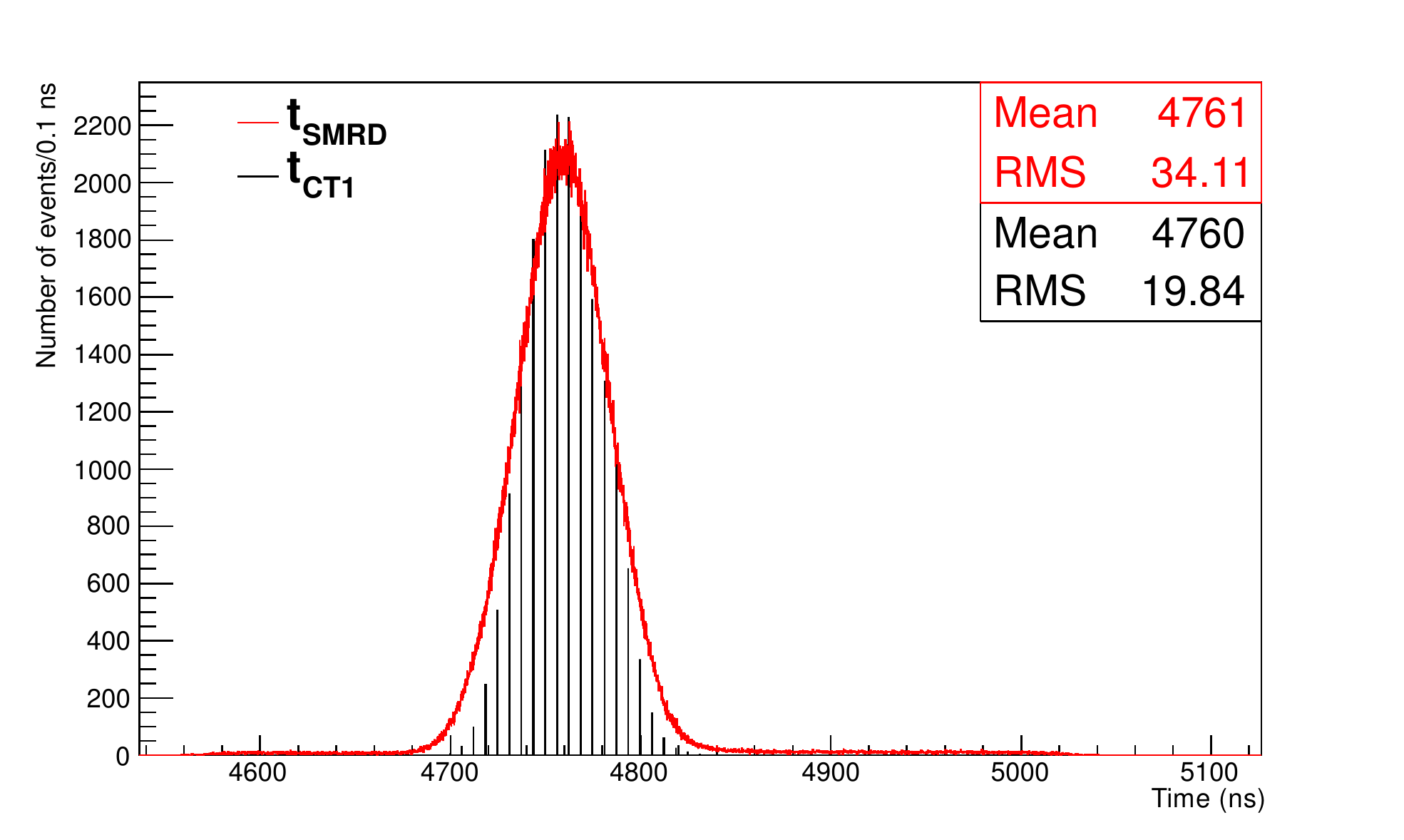} 
\caption{\label{fig:1bunch} Zoomed in version of the fourth bunch of the hits selected in SMRD (red: t$_{SMRD}$) and CT1 (blue: t$_{CT1}$) for run period III. }
\end{figure}

Figure~\ref{fig:8bunch} shows the resulting bunch timing distribution for the selected SMRD hits integrated over run period III. For comparison the bunch timing structure as observed with the current transformer CT1 also integrated over run period III is overlaid with appropriate delays included. The plotted timing distributions are for t$_{SMRD}$ and t$_{CT1}$, quantities which are described in more detail later.
Figure~\ref{fig:1bunch} shows a zoomed version of the fourth bunch for run period III and is representative for any other of the eight bunches. The peak heights of the two distributions have been scaled to match each other in order to facilitate a comparison of the bunch widths. The shown distributions include fluctuations in the inter-bunch timing from one spill to the next.

The CT1 data are recorded with a 160 MHz FADC which limits the timing resolution (e.g. time bins of 6.25 ns width). Timing jitter of the FADC start gate does not contribute significantly to the CT1 timing resolution. 

Selected hits in the SMRD originate from a combination of neutrino interactions in the iron of the magnet yokes and the surrounding sand. The background contribution from random noise hits per bunch is estimated to be well below 0.9\%. Additional components of  the sub-percent level background are due to decay electrons from muons stopping in the SMRD and ejected neutrons produced in the interactions of the proton beam with the target. 

The majority of these background events are removed by the requirement that events fall within a 200~ns time window centered on the peak of the SMRD bunch. The exact window size varies between bunches and runs and is determined based on the full CT1 bunch width plus a margin of 40~ns to allow for differences between the proton and neutrino bunch widths as measured by CT1 and the SMRD, respectively.
A cross-check with neutrino candidate events in the P$\O$D shows that this requirement does not lead to a reduction of neutrino candidate events in the sample.

Signal propagation time differences arising from the varying counter locations which are known to within 1~cm and readout cable lengths which are known to better than a cm have been corrected. Any remaining differences due to these corrections are well below the 1 ns level and are negligible.

\begin{figure}[thb]
\includegraphics[width=0.5\textwidth]{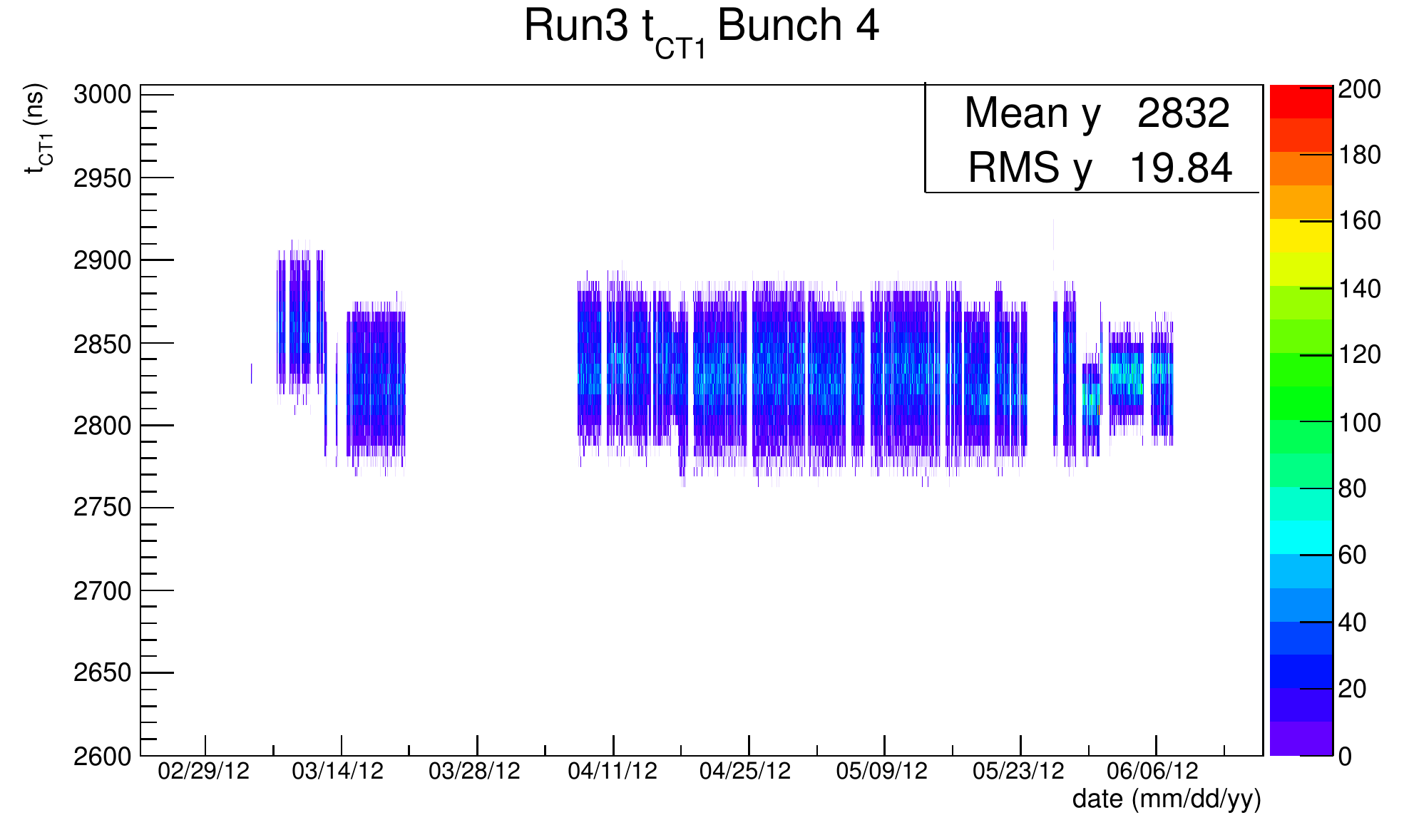}
\caption{\label{fig:run3CT1B4}Data for t$_{CT1}$ as function of calendar time for run period III and bunch 4.}
\end{figure}

\begin{figure}[h]
\includegraphics[width=0.5\textwidth]{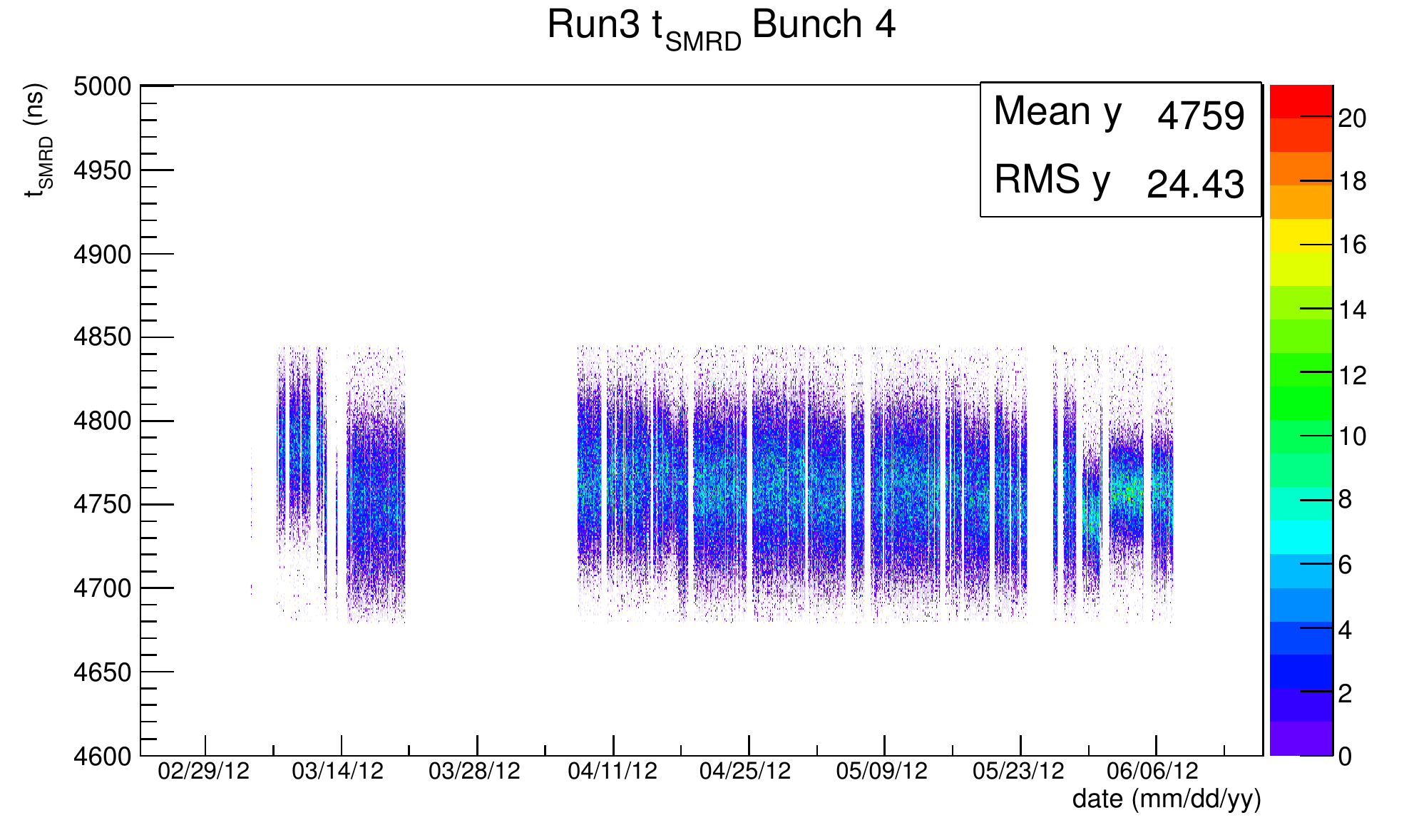}
\caption{\label{fig:run3SMRDB4}Data for t$_{SMRD}$ as function of calendar time for run period III and bunch 4.}
\end{figure}

\begin{figure}[h]
\includegraphics[width=0.5\textwidth]{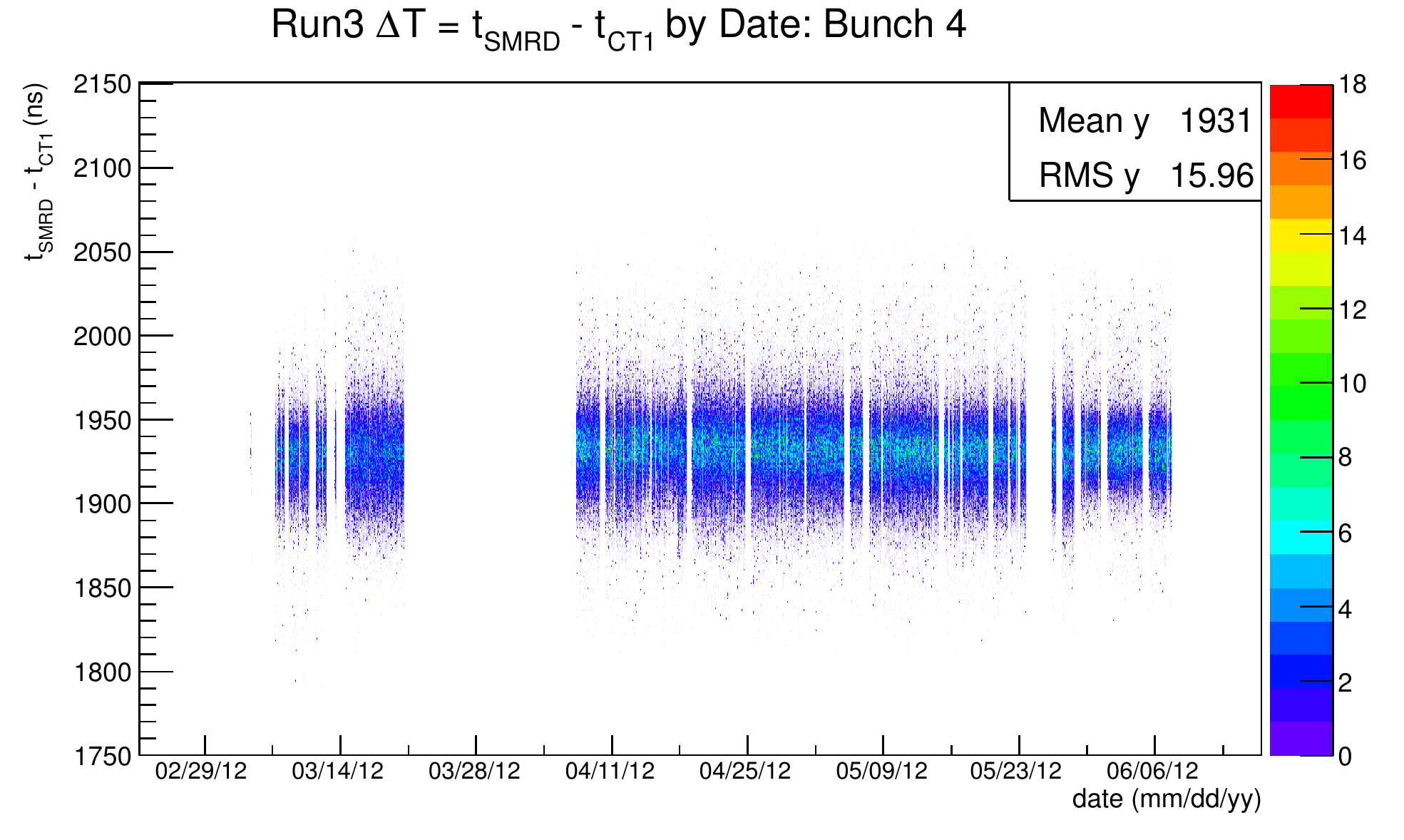}
\caption{\label{fig:run3CTSMRDB4}Data for $\Delta T = t_{SMRD} - t_{CT1}$ as function of calendar time for run period III and bunch 4.}
\end{figure}

An important difference in the measurement of the bunch timing by CT1 and the SMRD is that CT1 measures the proton beam signal and hence observes a signal which stems from all protons in a bunch for every single bunch. In contrast the SMRD observes muons from a single or a few neutrino interactions per bunch and thus sees the neutrino beam bunches. The distribution of CT1 timing signals t$_{CT1}$ is a direct measurement of the bunch to bunch arrival time fluctuations convoluted with the CT1 timing resolution.
Figure ~\ref{fig:run3CT1B4} shows distributions of t$_{CT1}$ as function of calendar time for bunch 4 and run period III.\\
The distribution of SMRD signal times t$_{SMRD}$  is a combination of the bunch width, variations in the bunch width, bunch to bunch arrival time fluctuations and the SMRD timing resolution for the selected event sample. 
Figure~\ref{fig:run3SMRDB4} shows distributions of t$_{SMRD}$ as function of calendar time for bunch 4 and run period III.\\
Figures~\ref{fig:run3CT1B4} and \ref{fig:run3SMRDB4} show times to have RMS values of 20~ns and 24~ns  over the run period. 
Beam event time measurements at CT1 and SMRD are affected by fluctuations in the arrival time $\sigma_{T_0}$ of the beam bunch. Such fluctuations originate from the accelerator complex and since CT1 and the SMRD use the same time stamped FX beam trigger signal, these fluctuations are common to both the CT1 and SMRD measurements and can be subtracted out. By looking at the difference $\Delta$T = t$_{SMRD}$  - t$_{CT1}$ on a bunch by bunch basis common fluctuations in the bunch arrival times are removed.
Figure~\ref{fig:run3CTSMRDB4} shows distributions of $\Delta$T as function of calendar time for bunch 4 and run period III and exhibits an RMS value of 16~ns over the run period.

The width $\sigma_{smrd+bunch}$  of the near detector timing probability distribution function (PDF) is calculated as 
\begin{eqnarray}
\sigma_{smrd+bunch}  = \sqrt{\frac{\sigma^2 _{SMRD} - \sigma^2 _{CT1} + \sigma^2 _{\Delta T}}{2}}
\end{eqnarray}
The three uncertainties $\sigma_{CT1}$ , $\sigma_{SMRD}$ and $\sigma_{\Delta T}$ can be directly determined from the SMRD and CT1 data sets
as shown in Figs.~\ref{fig:run3CT1B4}, \ref{fig:run3SMRDB4} and \ref{fig:run3CTSMRDB4}.

\begin{figure}[h]
\includegraphics[width=0.5\textwidth]{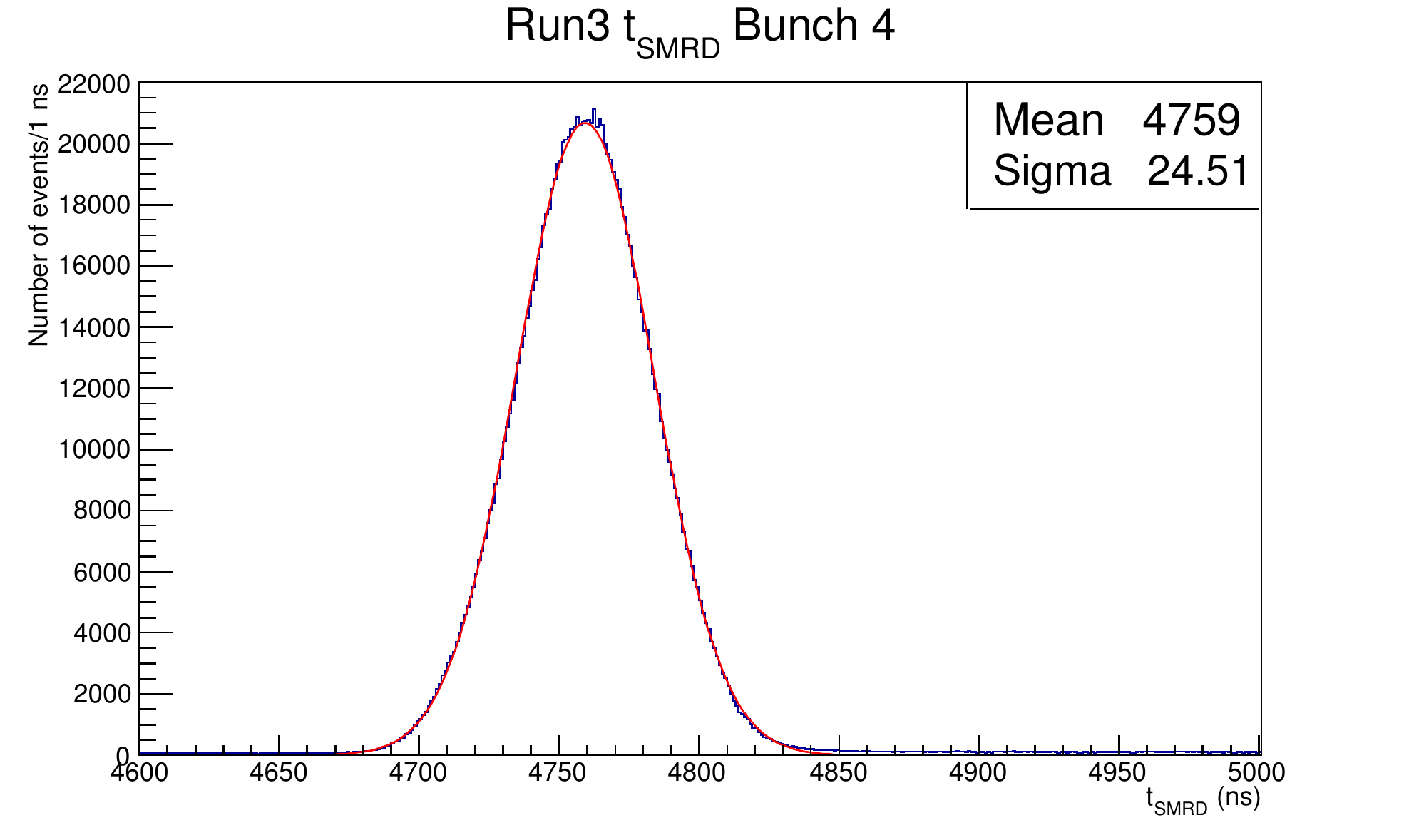}
\caption{\label{fig:run3SMRDB4_int} Distribution of $t_{SMRD}$ for run period III and bunch 4 with a superimposed Gaussian fit.}
\end{figure}

The distributions are to very good approximation Gaussian and are used to derive the three uncertainties $\sigma_{CT1}$ , $\sigma_{SMRD}$ and $\sigma_{\Delta T}$
on a run period by run period basis. 
An example of the time integrated distribution of SMRD signal times t$_{SMRD}$ for the 4th bunch and run period III  is shown in Fig.~\ref{fig:run3SMRDB4_int}.

Table~\ref{tab:SMRD_resol_runs} provides a summary of the SMRD hit timing resolutions, $\sigma_{smrd+bunch}$ for run periods I through IV and after averaging over all 8 (6 for run period I) bunches.
\begin{table} 
\caption{\label{tab:SMRD_resol_runs} SMRD hit timing resolution, $\sigma_{smrd+bunch}$ for run periods I through IV. The uncertainties represent the standard deviation between bunches 1 through 8. Run period II data is split into two parts due to a break in data-taking and associated system resets. }
\begin{ruledtabular}
\begin{tabular}{c c}
Run period number & SMRD hit timing resolution \\
   &  $\sigma_{smrd+bunch}$ [ns] \\ \hline
I  & 12.1 $\pm$ 0.1 \\
IIa  &  12.6 $\pm$ 1.7 \\
IIb  & 14.8  $\pm$ 0.5 \\
III  & 14.5 $\pm$ 0.3 \\
IV  &  13.8 $\pm$ 0.8 \\
\end{tabular}
\end{ruledtabular}
\end{table}

A more detailed list of resolutions 
$\sigma_{CT1}$ , $\sigma_{SMRD}$, $\sigma_{\Delta T}$ and $\sigma_{smrd+bunch}$
is summarized in table~\ref{tab:SMRD_resol_runs_detail} in the appendix for each of the 8 (or 6) bunches and all four run periods.

\subsection{Far detector data selection and timing}

At the far detector FC events and FCFV single ring muon-like and electron-like events 
all of which are in time with beam spills are selected. Additional selection cuts on the charged lepton momentum and the 
presence or lack of decay electrons from stopping muons for the muon-like and electron-like single ring event sample are applied
to increase the purity of CCQE events in the sample. The electron-like single ring sample has further selection cuts to reduce contamination from 
$\pi^0$ and other backgrounds.

Details of the selection criteria for FC and CCQE
candidate events are described in \cite{T2K_results} and are the same as used for T2K oscillation analyses.
A total of 549 FC events and 148 CCQE candidate events in time with beam spills were
observed at SK during run periods I through IV. 

\begin{figure}
	\begin{center}
	\subfigure{
		\includegraphics[width=0.45\textwidth]{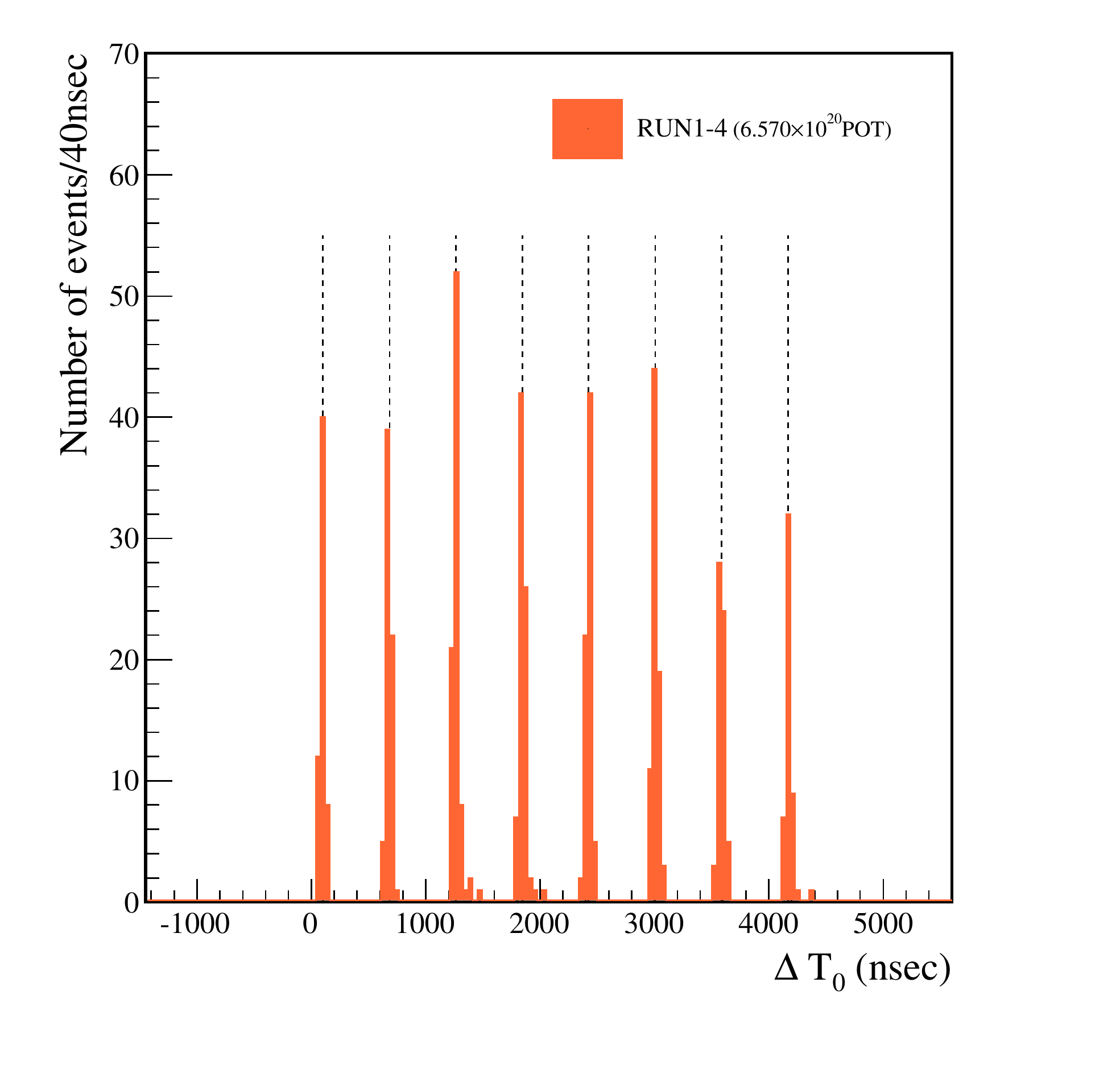} 
	}		
	\subfigure{
		\includegraphics[width=0.45\textwidth]{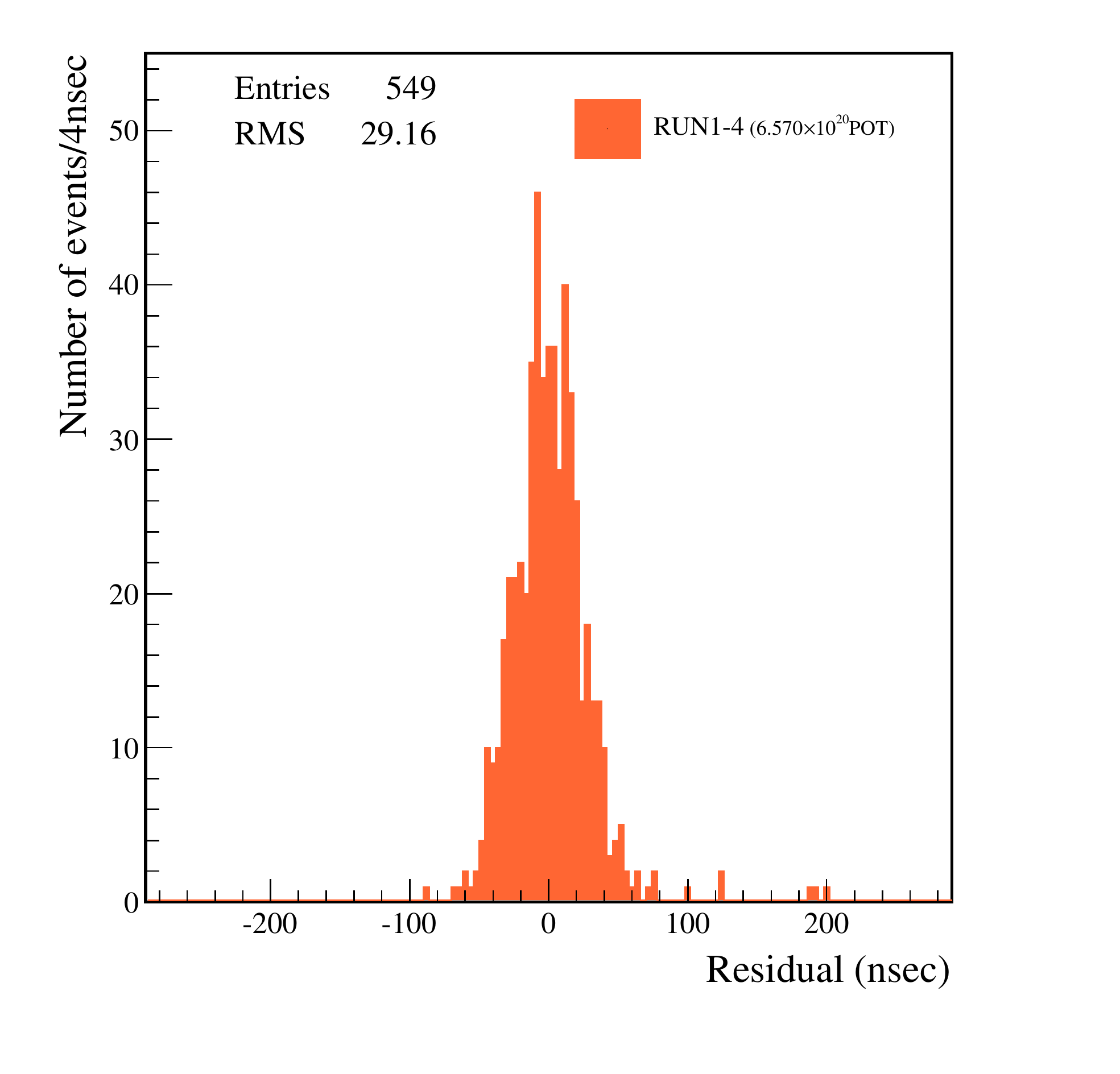} 
	}		
	\end{center}
\caption{\label{fig:SK_time_res} Timing distribution (top) and timing residuals (bottom) of fully contained (FC) events for run periods I $-$ IV.}
\end{figure}

The resulting timing distribution for FC events is shown in the upper panel of Fig.~\ref{fig:SK_time_res}, which clearly reflects the beam bunch structure. 
Event times are corrected for differences in the reconstructed event vertex positions and associated neutrino and light travel times. The eight dotted vertical lines in Fig.~\ref{fig:SK_time_res} represent the bunch center positions fitted to the observed FC event timing peaks preserving the 581ns inter-bunch intervals. 
The distribution of the residual time between each FC event and the closest fitted bunch center is shown in the lower plot of Fig.~\ref{fig:SK_time_res} 
for events combined from all bunches and all 4 run periods. The outliers at residual values of +200~ns are classified as decay electrons based on predetermined criteria including event single ring properties, electron like particle identification and visible energy.

The SK event selection relies on the relative GPS system timing between the near and far sites. Hence, the distribution of SK event times $t_{SK}$ and time residuals with respect to the nearest beam bunch center include instabilities associated with the GPS timing system at the near and far sites. The distribution of SK signal times $t_{SK}$  is a combination of the bunch width, variations in the bunch width,
the SK timing resolution for the selected event sample, GPS system timing instabilities and potential neutrino TOF effects.

Our data analysis uses the distribution of timing residuals of FC events excluding the CCQE candidate event sample (FC non-CCQE)
 \cite{SK} with a visible energy above 2 GeV on a run period by run period basis to determine the combined SK and GPS system timing resolution. 
This sample is orthogonal to the CCQE candidate sample used to extract a neutrino mass limit and will therefore be referred to as 
sideband sample FC$_{sideband}$.
The cut on visible energy was introduced so as to avoid potential bias of the timing resolution due to time of flight related relativistic delays in neutrino arrival times. Above a visible energy of about 2 GeV and for neutrino masses m$_{\nu} < $6~MeV$/c^2$ relativistic delays are expected to be less than about 5 ns. The uncertainties on the combined SK and GPS resolutions are of comparable size, 4 to 5ns.

\begin{figure}
\includegraphics[width=0.5\textwidth]{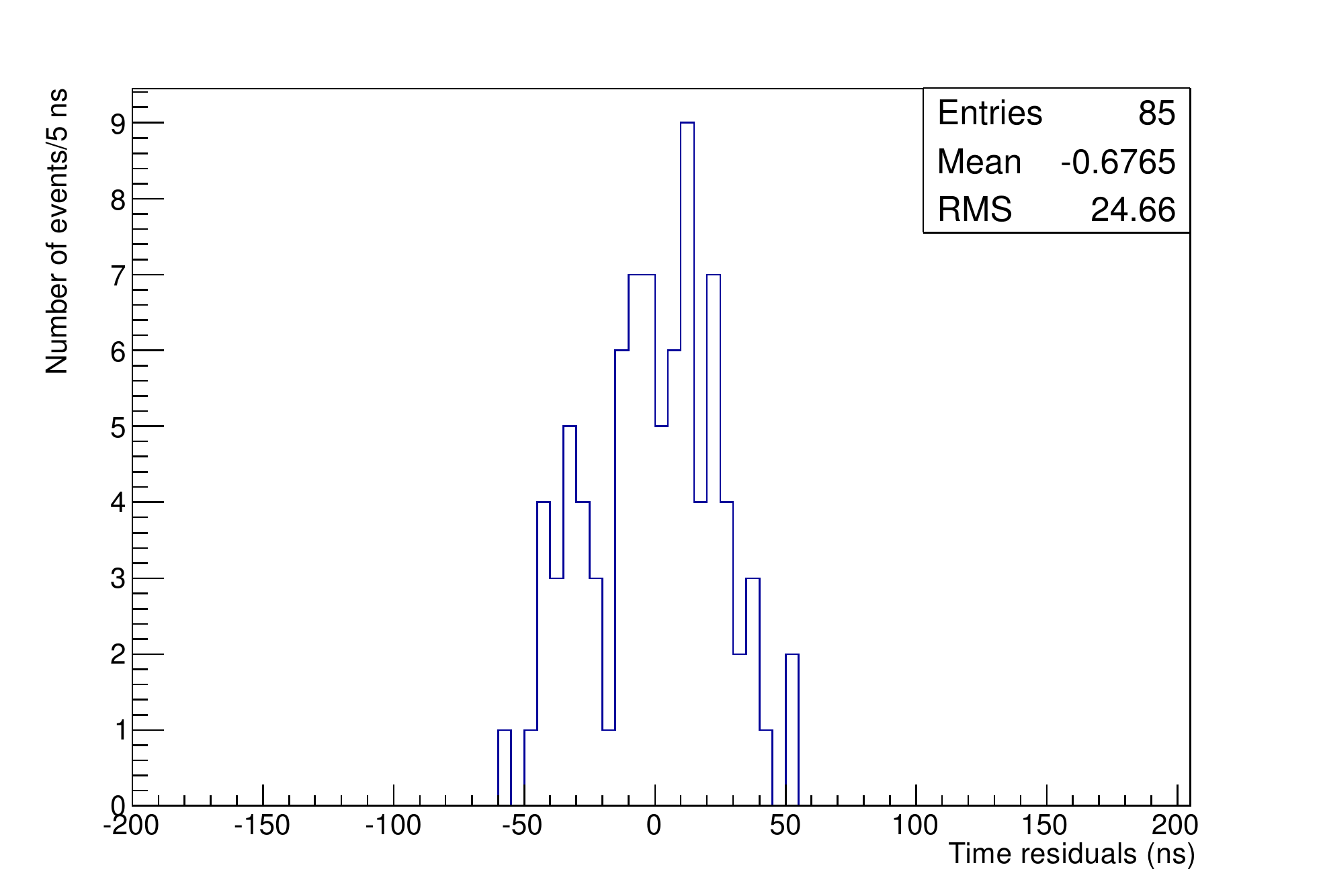} 
\caption{\label{fig:nonCCQE_run3} SK residual timing distributions for run period III for FC non-CCQE sample.}
\end{figure}

Figure~\ref{fig:nonCCQE_run3} shows the distribution of timing residuals for FC non-CCQE events.
Table~\ref{tab:SK_RMS_res} provides a summary of the measured combination of SK and GPS system timing resolutions and the bunch width
for run periods I through IV after averaging over individual bunches. Shown are the RMS values of the timing residuals distributions for each run period. Due to limited statistics for run period I the RMS value was derived for events from run period I and II combined. The errors are statistical and calculated based on the number of events in the corresponding run period. 
\begin{table} 
\caption{\label{tab:SK_RMS_res} Measured combination of SK and GPS timing resolutions and bunch width obtained from FC$_{sideband}$ events 
for run periods I through IV. }
\begin{ruledtabular}
\begin{tabular}{c c}
Run period number & RMS of SK FC$_{sideband}$ event \\
   &  timing residulas [ns] \\ \hline
I + II (combined)  & 19.4 $\pm$ 4.3 \\
II &  21.4 $\pm$ 5.2 \\
III  & 22.9 $\pm$ 5.9 \\
IV  &  26.0 $\pm$ 3.6 \\
\end{tabular}
\end{ruledtabular}
\end{table}

\begin{figure}
\includegraphics[width=0.5\textwidth]{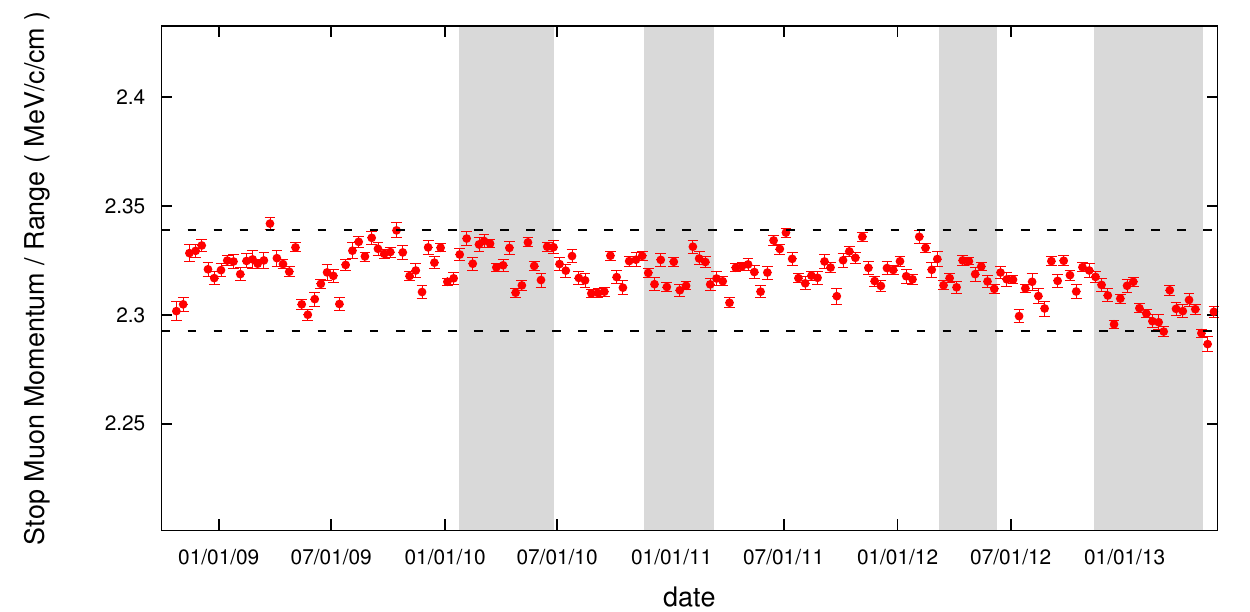} 
\caption{\label{fig:SK_stopmu_mom_stability} Energy scale stability of the SK detector \cite{SK}. The hatched grey regions correspond to run periods I through IV. The dotted horizontal lines correspond to the $\pm$1\% stability range.}
\end{figure}

The stability of the SK energy scale over the relevant data period is demonstrated in Fig.~\ref{fig:SK_stopmu_mom_stability}.
It shows the observed energy loss for stopped muons as a function of calendar time. The energy scale is stable within $\pm$1\% over the run period I to IV time range.  Reference \cite{SK} also specifies the total error on the energy scale to be 2.3\% for run periods I through III and 2.4\% for run period IV.

\subsection{GPS system timing}

\begin{figure}[ht]
	\begin{center}
	\subfigure{
		\includegraphics[width=0.45\textwidth]{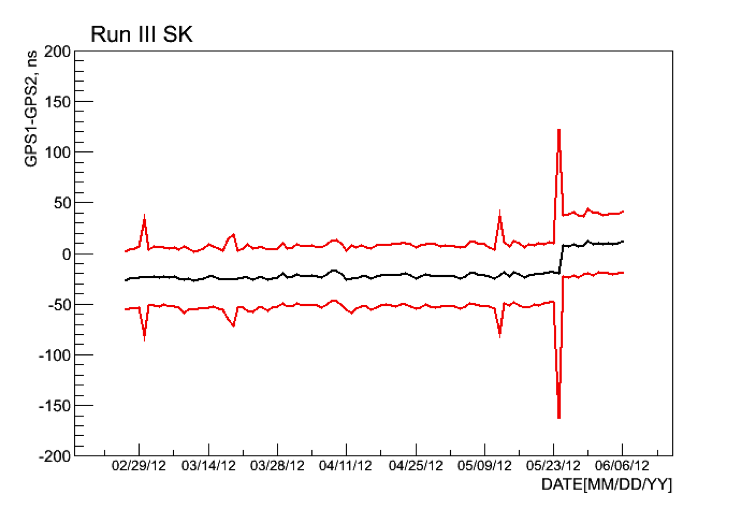} 
	}		
	\subfigure{
		\includegraphics[width=0.45\textwidth]{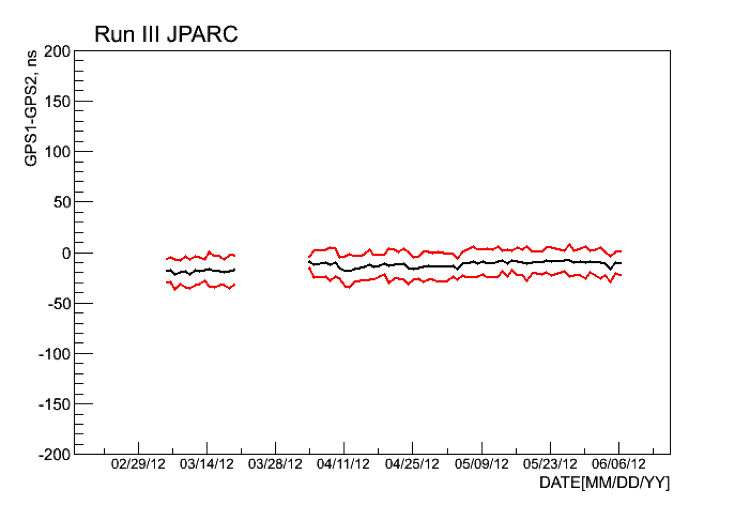} 
	}		
	\subfigure{
		\includegraphics[width=0.45\textwidth]{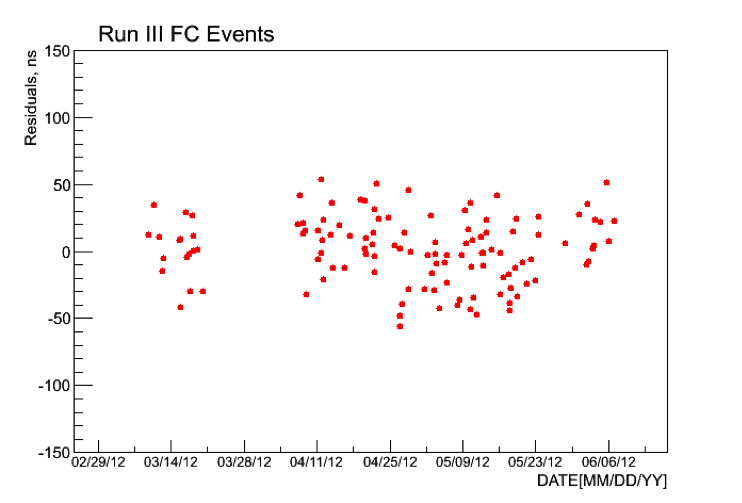} 
	}		
	\end{center}
\caption{\label{fig:GPS-stability}Relative timing stability of the two official SK GPS systems at SK (top panel) and NU1 (middle panel). Black and red lines are daily averages and  95 \%C.L. ranges, respectively. The bottom panel shows the time residuals of the SK FC events. 
}
\end{figure}

The quality of the relative timing between the SK and NU1 GPS reference times relies on the stability of the GPS time systems at SK and NU1. 
 The time differences between GPS1 and GPS2 receivers at SK and NU1 are shown in the top and middle panels of Fig.~\ref{fig:GPS-stability} as a function of time for run period III. 
The center line in each plot shows the 24 hour average difference between primary and secondary GPS at each site. The daily upper and lower 95\% C.L. fluctuation ranges are also shown by red lines. Occasional abrupt baseline shifts with magnitudes of several tens of ns are typically a consequence of a system reset of either one of the GPS receivers,
causing an oscillation shift upon restart. 
Jumps during a given  run have been analyzed in detail to check whether neutrino candidate event timing data would be affected.
 Since data from each run period are analyzed separately the shifted baselines between run periods do not affect the analysis results. 
 The bottom panel of Fig.~\ref{fig:GPS-stability} shows the time residuals of FC events at SK with respect to the time of the closest beam bunch center. 

During run period III the observed step on 5/25/2012 in the GPS1$-$GPS2 time history at SK is due to a reset of GPS2. At the time of the incident GPS1 was not affected. Since candidate events at SK obtain time stamps from GPS2 the time stamps of FC events post the GPS2 reset on 5/25/2012 have been corrected. The correction consists of a time shift of the affected events by the magnitude of the observed step.  Shifts of order of few tens of nanoseconds in baseline after a GPS reset are expected.
In addition, instantaneous GPS status information is logged and has been used to further scrutinize the quality of time stamps for the recorded FC candidate events.
All time stamps for the FC events can be considered reliable.
 
The timing uncertainty derived from the SK FC$_{sideband}$ sample already contains any timing instabilities due to the GPS systems at the near and far site since the SK event selection relies on the time stamps at the near and far sites.

\section{DATA ANALYSIS}
\label{sec:DataAnal}
\label{norm_offset}
The same neutrinos cannot be observed at the near and far detector. However, we can use the shape of the beam timing structure at the near detector and neutrino candidate events at the far detector to obtain a time of flight measurement. We construct a suitable probability density function (PDF) that describes the expected arrival times of the neutrino beam bunch structure at the far detector and fit it to the actual distribution of neutrino candidate events at SK. Differences between expected and actual arrival time distributions can be determined and used to derive neutrino RTOF differences between events. The differences are studied as a function of neutrino event energy and therefore can be used to set an upper limit on neutrino mass.

The total resolution for neutrino arriveal times at SK will be affected by the detector timing resolution and jitter from the GPS timing system. The shapes of the bunches in the near detector are to very good approximation Gaussian as can be seen in Fig.~\ref{fig:1bunch}.
Hence the near detector PDF is represented by 6 or 8 Gaussian distributions each with a width of, $\sigma_{smrd+bunch}$ which reflects a combination of the SMRD timing resolution and the bunch width. Values of $\sigma_{smrd+bunch}$ for each run period are given in table~\ref{tab:SMRD_resol_runs}.  
The far detector PDF is constructed by convoluting the near detector PDF with a Gaussian. 
For the width of the Gaussian we use the measured values for $\sigma_{SK+GPS}$, which represent an upper bound on the combined SK and GPS system timing resolutions.
The far detector PDF can be written as:
\begin{eqnarray}
P_2 (t_2) = \int \frac{1}{\sigma_{SK+GPS} \sqrt{2 \pi}} \, e ^{-\left(\frac{(t_2-t)^2}{2 \sigma^2_{SK+GPS}}\right)} P_1(t) dt 
\end{eqnarray}
where $\sigma_{SK+GPS}$ is the timing uncertainty due to SK and the GPS system, P1 is the near detector PDF for any of the 6 or 8 bunches, and $t_2$ is the measured time of FC CCQE events in the far detector. 
The resulting distribution describes the expected arrival times at SK. The combined SK, GPS and SMRD hit timing resolution which is reflected in the width of the bunches of $P_2$ was determined according to
\begin{eqnarray}
\sigma_{P_2} = \sqrt{\sigma^2_{SK+GPS} + \sigma^2_{smrd+bunch}}
\end{eqnarray}
and found to be about 27 ns for run periods I through IV. Relevant values for $\sigma_{smrd+bunch}$ and $\sigma_{SK+GPS}$ are presented in 
tables~\ref{tab:SMRD_resol_runs} and \ref{tab:SK_RMS_res}, respectively.

For a given run period the times of all CCQE events are adjusted by a common off-set. The off-set is chosen such that the mean time for all CCQE candidate events with a derived neutrino energy above 2 GeV is zero for that run period. The arrival time distribution of high energy events is not sensitive to the neutrino mass, m$_{\nu}^2$, for m$_{\nu}^2 <$ 10 MeV$^2/c^4$. Hence these events represent a good sub-sample for adjusting all events. The adjustment off-sets relative to the mean arrival time of the 
FC sample for run periods I+II combined, III and IV are -2.5 ns, 1.1 ns and 5.0 ns, respectively.

We compare the timing of each event in the far detector to this PDF and find the value m$_{\nu}^2$ by minimizing a negative log-likelihood function L over the CCQE neutrino candidate events in the sample.
\begin{eqnarray}
L = \sum_i - ln P_2(t_2^i - T_{m_{\nu}}(E_{\nu}) )
\end{eqnarray}
where $t_2^i$ are the residual times of neutrino candidate events and
T$_{m_{\nu}}(E_{\nu})$ represents the relativistic time of flight of massive neutrinos and is given by
\begin{eqnarray}
T_{m_{\nu}}(E_{\nu}) = \frac{\tau}{\sqrt{1- ( \frac{m_{\nu} c^2}{E_{\nu}})^2}} \,\,,
\end{eqnarray}
with $\tau$ being the light travel time.

The previously mentioned adjustment of measured event timing residuals for events with derived neutrino energies E$_{\nu} >$ 2 GeV and m$_{\nu}^2 < $10  MeV$^2/c^4$ can be expressed as
\begin{eqnarray}
\label{eqn:normalization}
\overline{( t_2^{i} - T_{m_{\nu}}(E_{\nu}))} \,\,  \approx \,\, \overline{(t_2^{i} - \tau)} =0 
\end{eqnarray}m
The neutrino energy E$_{\nu}$ is derived from the reconstructed energy E$_{\nu -recon}$ of the detected event, which is calculated, neglecting the Fermi motion, as
 \begin{equation}\label{eqn:erecccqe}
E_{\nu -recon}= {{m^2_p-(m_{n}-E_b)^2-m^2_{\mu}+ 2 (m_n -E_b) E_{\mu} } \over {2(m_{n}-E_b-E_{\mu}+p_{\mu}\cos \theta_{\mu})}},
\end{equation}
where $m_p$ is the proton mass, $m_{n}$ the neutron mass, and $E_b=27$~MeV
the binding energy of a nucleon inside a $^{16}$O nucleus. In Eq.~\ref{eqn:erecccqe} $E_{\mu}$, $p_{\mu}$, 
and $\theta_{\mu}$ are respectively the measured muon energy, momentum and angle with respect to the incoming neutrino.

\begin{figure}[h]
\includegraphics[width=0.48\textwidth]{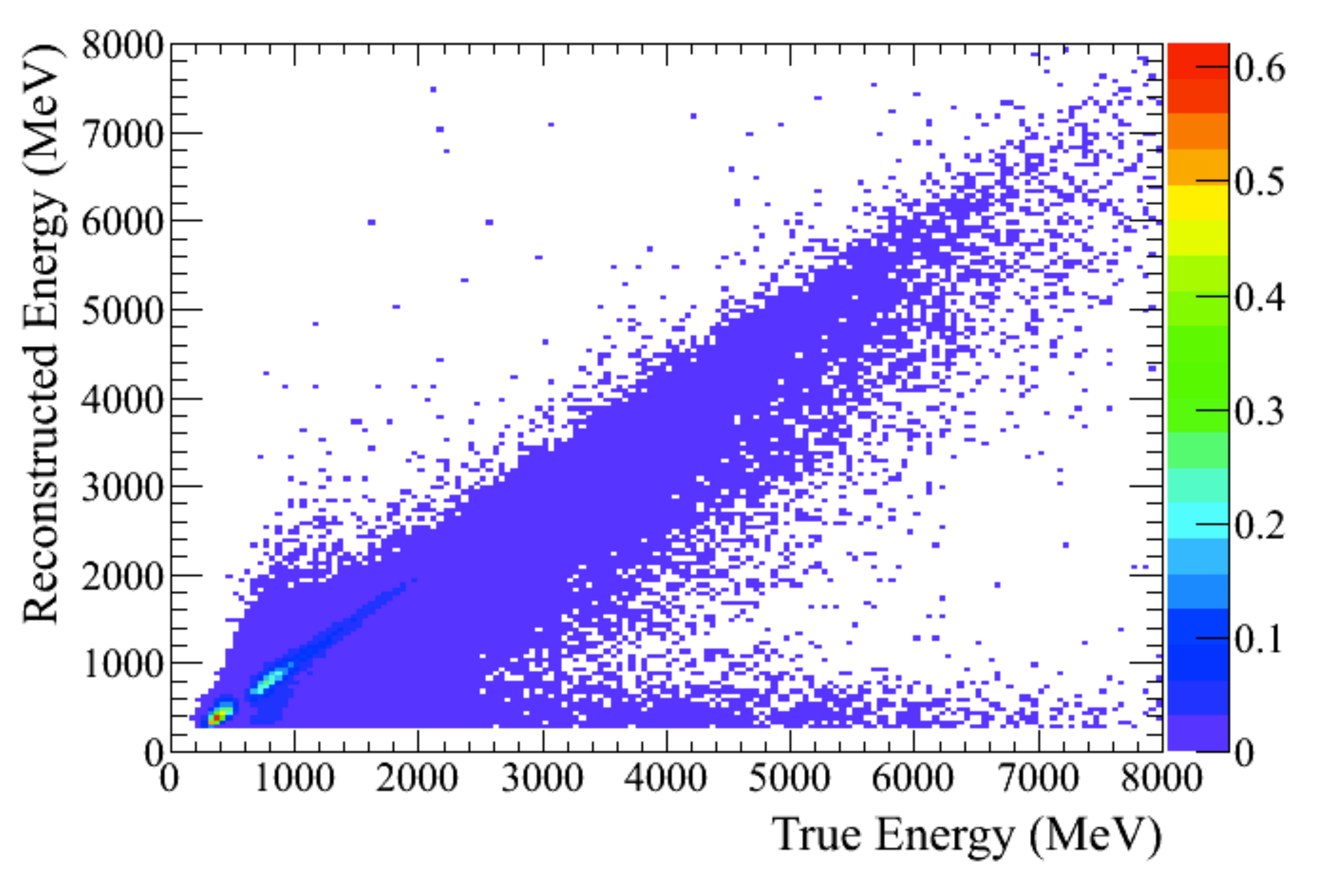} 
\caption{\label{fig:conversion_matrix} Monte Carlo energy conversion matrix for muon neutrino candidate events. The vertical axis shows reconstructed neutrino energy and the horizontal axis shows true neutrino energy. }
\end{figure}

A detector response matrix relating true neutrino energy and reconstructed neutrino energy has been obtained from the SK Monte Carlo (MC). 
Selection cuts to identify FCFV single ring muon-like and electron-like events are applied to the MC events to derive a separate matrix for each selection \cite{SK,T2K}. The MC data samples are composed of CC and NC interactions for initial muon neutrinos, muon anti-neutrinos, electron neutrinos and electron anti-neutrinos with relative percentages of 92.8\%, 6.0\%, 1.0\% and 0.2\%, respectively \cite{T2K_flux}. Neutrino oscillation effects are included in the MC data samples by re-weighting energy spectra with oscillation probabilities. Table~\ref{tab:oscil_paras} shows the values of the assumed oscillation parameters.
\begin{table} 
\caption{\label{tab:oscil_paras} Oscillation parameters to calculate oscillation probabilities for purposes of reweighing energy spectra }
\begin{ruledtabular}
\begin{tabular}{c c}
Oscillation parameter & value \,\,\,\,\,\\ \hline
$\Delta m^2_{23}$ & $2.4 \times 10^{-3}$ \,\,\,\,\, \\ 
$\sin^2 (2 \theta_{23})$ & 1.0 \,\,\,\,\, \\ 
$\sin^2 (2 \theta_{13})$ & 0.1 \,\,\,\,\, \\ 
$\sin^2 (2 \theta_{12})$ & 0.87 \,\,\,\,\,\\ 
$\Delta m^2_{12}$ & $7.6 \times 10^{-5}$ \,\,\,\,\, \\ 
\end{tabular}
\end{ruledtabular}
\end{table}
For a given data set of CCQE candidate events, each with a reconstructed energy, true energies are obtained by sampling from the corresponding entries in  the conversion matrix which is shown in Figure~\ref{fig:conversion_matrix}.
From the resulting data samples with true neutrino event energies an average negative log-likelihood is calculated. In the minimization the neutrino mass squared m$_{\nu}^2$
is left free to vary. The average log-likelihood yields a best-fit m$_{\nu}^2$ value and a negative log-likelihood curve as function of m$_{\nu}^2$.
The validity of the analysis method is tested on an ensemble of 300 toy data sets and is described in the following section. The same statistical sampling procedure of the true energy distribution is also applied to the experimentally recorded data set.

\subsection{Toy data studies}

The performance of the analysis was tested successfully on an ensemble of 300 toy data sets which are based on T2K Monte Carlo data samples for the SK detector. Each toy data set consists of 148 CCQE candidate events and is subdivided into 4 subsets, corresponding to the number of events observed in run period I through IV.
 The same FCFV 1 ring muon-like and electron-like event selection cuts were applied to the toy data sets as for the experimentally recorded T2K-SK data. 
 Each of the resulting candidate events was assigned a residual bunch time.  This residual time was sampled from a Gaussian distribution whose width was determined according to the SK FC$_{sideband}$ event timing residuals for the corresponding run periods. The numerical values of the widths used are shown in table~\ref{tab:SK_RMS_res}. Initially, all of the 300 toy data sets assume a true value of m$^2_{\nu-true}$ = 0.
	Each toy data set was submitted to the previously described analysis procedure. 

Further tests of the analysis algorithm were conducted to study the sensitivity of the analysis to large changes of the assumed true m$^2_{\nu-true}$. 
The tests were performed successfully on an ensemble of 300 toy MC data sets with 148 events each. 
The toy data sets are similar to previously described toy data sets except for modified event times to reflect the effect of non-zero values of m$^2_{\nu-true}$. 
Event time residuals were modified according to an assumed neutrino mass squared of  m$^2_{\nu-true}$ = 1, 2, 4, 10, 20 and up to 100 MeV$^2/c^4$  in steps of 10 MeV$^2/c^4$ . The modified event times are calculated according to equation~\ref{eqn_time}:
\begin{eqnarray}
\label{eqn_time}
t = \tau - \frac{\tau}{\sqrt{1- ( \frac{m_{\nu} c^2}{E_{\nu}})^2}}
\end{eqnarray}

\begin{figure}
\includegraphics[width=0.5\textwidth]{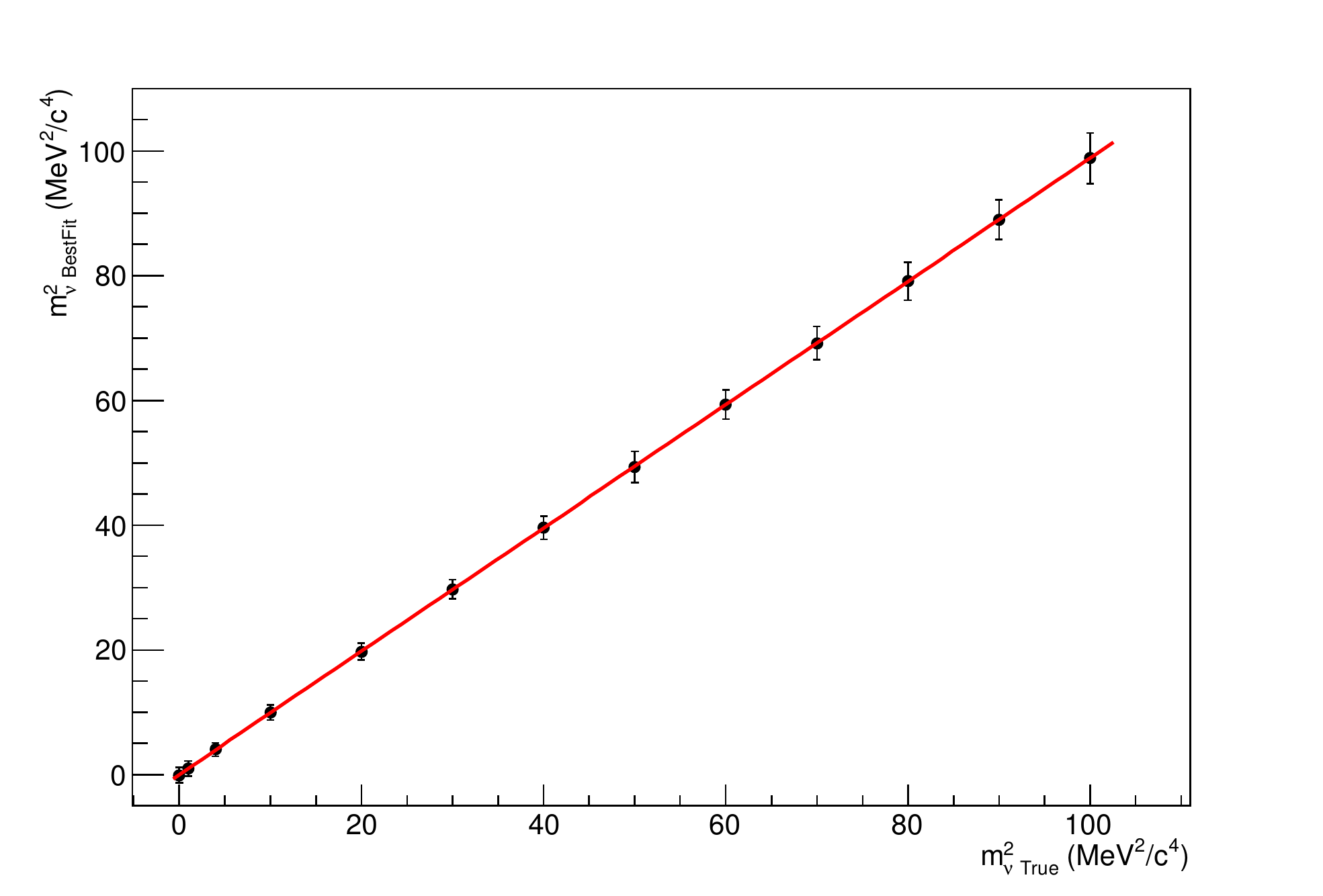} 
\caption{\label{fig:toy_mnu_vs_true} Central value of distributions of best fit values of m$^2_{\nu}$ best-fit versus m$^2_{\nu-true}$  for 300 toy MC data sets for which event arrival times had been shifted based on the assumption of the shown m$^2_{\nu-true}$ values. The error bars represent the RMS of the corresponding distributions.}
\end{figure}

Figure~\ref{fig:toy_mnu_vs_true} shows the extracted values for m$^2_{\nu}$ versus the corresponding true input values m$^2_{\nu-true}$ for the full range of tested 
m$^2_{\nu-true}$ input values. The error bars represent the RMS of the distribution of the 300 extracted best-fit values for m$^2_{\nu}$.
The analysis correctly obtained the true input values for m$^2_{\nu}$.

 \subsection{T2K data analysis}

The previously described analysis procedure of fitting a timing PDF to experimentally recorded timing distributions is applied to the SK event sample of CCQE neutrino candidate events. Data from run periods I through IV are fitted by setting up a single log-likelihood function with run specific PDFs. 
This final signal data set which is used to extract a limit on the effective neutrino mass contains 120 muon neutrino and 28 electron neutrino CCQE candidate events. For this data set the true event energies were sampled 1000 times from the true energy distributions for the reconstructed energy of each event. 

\begin{figure}
\includegraphics[width=0.5\textwidth]{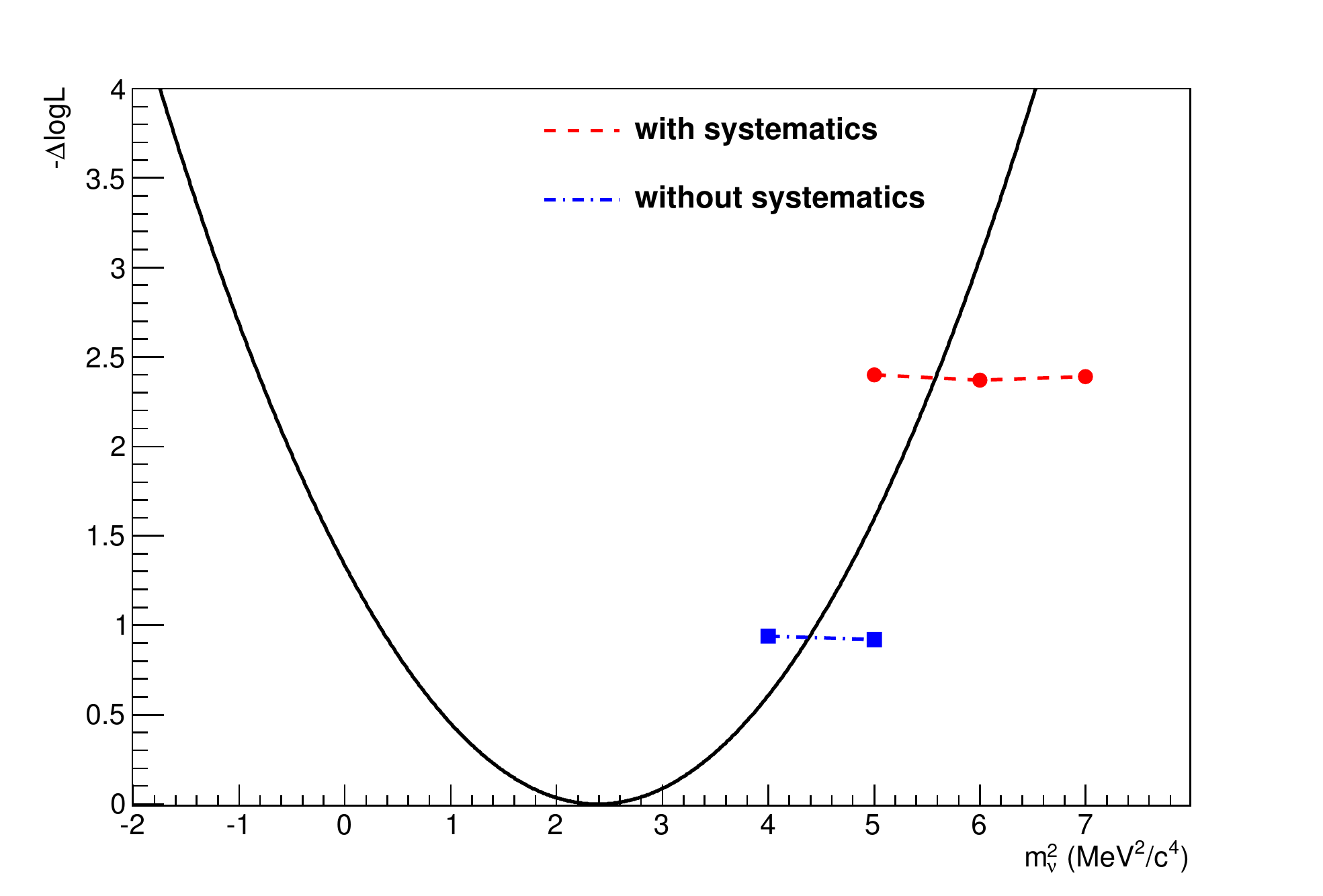} 
\caption{\label{fig:CCQE_av_logL} Log-likelihood curve averaged over 1000 sampled true neutrino energies from the energy conversion matrix. The minimum value is found to be m$^2_{\nu}$ = 2.4 MeV$^2/c^4$. The blue squares and red dots indicate the F-C 90\% C.L. upper critical values without and with systematic uncertainties, respectively.
The corresponding 90\% C.L. upper limits on  m$^2_{\nu}$ = 4.4 MeV$^2/c^4$ and m$^2_{\nu}$ = 5.6 MeV$^2/c^4$ are derived . }
\end{figure}

The log-likelihood curves for the 1000 energy samples are averaged to provide an average log likelihood curve, which is shown in Fig.~\ref{fig:CCQE_av_logL}. The best fit m$^2_{\nu}$ value is found to be at m$^2_{\nu}$ = 2.4 MeV$^2/c^4$. 
A 90\% C.L. upper limit is evaluated by means of an ensemble of toy data sets. We calculate a 90\% critical value following the Feldman-Cousins (F-C) method described in \cite{FC}. An ensemble of 300 toy data sets is generated with event times corresponding to specific value of m$^2_{\nu-true}$. For each of these toy data sets an average Delta-log-likelihood value at the given value of  m$^2_{\nu-true}$ is calculated. A distribution of these average Delta-log-likelihood values is integrated from 0 to 90\%. The Delta-log-likelihood value found at the 90\% integration boundary represents the critical value at that particular m$^2_{\nu-true}$. The process is repeated for additional values of 
m$^2_{\nu-true}$ and the resulting 90\% C.L. critical values are shown along with the average log-likelihood curve for the experimental data set. The intersection of these two curves is used to determine the 90\% C.L. upper limit on m$^2_{\nu}$.
Figure~\ref{fig:CCQE_av_logL} shows the F-C 90\% C.L. upper critical values for  statistical uncertainty only as blue squares for m$^2_{\nu-true}$ values
of 4 and 5 MeV$^2/c^4$ superimposed on the average negative log-likelihood curve extracted from the data set.
A 90\% C.L. upper limit on m$^2_{\nu}$ of $\sigma^{90\%C.L.}_{m^2_{\nu}}$ = 4.4 MeV$^2/c^4$ is obtained.  This limit does not yet include systematic errors.

\subsection{Systematic Uncertainties}

Systematic uncertainties, which affect the extracted value of m$^2_{\nu}$ are caused by the event sample timing adjustment as well as the time resolution at the far detector, the near detector and the GPS system. Other sources of systematic uncertainties stem from reconstructing the event vertex, lepton momentum and the lepton direction with respect to the neutrino beam at SK. 

The effect of systematic uncertainties has been evaluated by regenerating ensembles of toy data sets, which had the systematic parameters varied within their 1 sigma bounds. Successively, the 90\% C.L. critical values are recalculated and modified 90\% C.L. upper limits on m$^2_{\nu}$ are derived.

The systematic uncertainty on the event sample timing adjustment is assumed to be the uncertainty on the residual times of all CCQE candidate events with a derived energy above 2 GeV. It is calculated as the RMS spread of the data points above 2 GeV shown in Fig.~\ref{fig:CCQE_dT_Eder_bin}
 and divided by the square root of the number of events. Values for this uncertainty for the different run periods are shown in table~\ref{tab:syst_tab}. 
 The adjustment offsets are within the systematic uncertainty.

\begin{table} 
\caption{\label{tab:syst_tab}  Summary of systematic effects on m$^2_{\nu}$. The left most column specifies the type of systematic uncertainty, the central columns give values for the different run periods and the right columns show the resulting uncertainty on the 90\% C.L. upper limit of m$^2_{\nu}$ and the change compared to the no-systematics case. The systematic error on lepton angle is 1$^o$ for lepton momenta below 1.33 GeV$/c^2$ and 2$^o$ for lepton momenta above 1.33 GeV$/c^2$. Values are taken from table~\ref{tab:SMRD_resol_runs} and \ref{tab:SK_RMS_res}.}
\begin{ruledtabular}
\begin{tabular}{c c c c c c c c}
Systematic & \multicolumn{5}{c}{Magnitude [ns]} & \multicolumn{2}{c}{90\% C.L.} \\
 uncertainties & \multicolumn{5}{c}{Run periods} & absolute & \% \\
&  I & IIa & IIb & III & IV & [MeV$^2/c^4$] & change \\ \hline \hline
Time & 4.4 &  4.4 &  4.4  & 5.6  &  5.0  &  5.09 &  +15.9 \\
adjustment & & & & & & & \\ \hline
Lepton  & \multicolumn{4}{c}{2.3\%}   & 2.4\%   & 4.4 &   +0.2 \\
momentum + & & & & & & & \\
angular bias & \multicolumn{5}{c}{1(2)$^o$} & & \\ \hline
SK + GPS time  &  4.3 &   5.2 &  5.2  & 5.9 & 3.6 &  4.75 &  +8.2 \\
resolution & & & & & & & \\ \hline
SMRD time & 0.1 &  1.7 &  0.5 &  0.3 &  0.8 &  4.41 &  +0.5 \\
resolution & & & & & & & \\ \hline
Total  & & & & & &  5.58 &  +27.1 \\ 
\end{tabular}
\end{ruledtabular}
\end{table}

Values for the systematic uncertainty on lepton momentum and the angular uncertainty of the lepton direction are based on systematic uncertainties derived from the SK atmospheric data sample \cite{SK}. We adopted a 2.3\% systematic uncertainty on the lepton energy scale at SK for run periods I through III.
For run period IV the corresponding value is 2.4\%.

The effect of systematic errors in the SMRD and combined SK and GPS time resolutions on m$^2_{\nu}$ was accounted for by changing the widths of the underlying Gaussian distributions from which the event times were sampled. Values of 0.1 to 1.7~ns for the SMRD and 5.9~ns to 3.6~ns for SK + GPS 
are the run period specific errors as shown in table~\ref{tab:SMRD_resol_runs} and \ref{tab:SK_RMS_res}, respectively.

Systematic uncertainties for the energy conversion matrix stemming from uncertainties in oscillation parameters as well as nuclear interaction effects 
have been estimated to be negligible. 
The magnitude of nuclear effects for various final state interaction (FSI) parameters has been obtained from MC studies \cite{NEUT} in form of 
modified reconstructed neutrino energies.
For each event the modified reconstructed energy is converted into a neutrino energy by sampling a 1000 times from the corresponding row in the conversion matrix. The resulting distributions of derived energies for FSI reweighted E$_{\nu}^{recon}$ spectra for toy MC data sets are consistent with the original (non-FSI reweighted) distributions of derived neutrino energies.  
 
The effect of the systematic uncertainties on m$^2_{\nu}$ were evaluated by repeating the previously described analysis with one type of systematic parameter varied
within its bounds and for all 4 subsets. 
Table~\ref{tab:syst_tab} summarizes the systematic uncertainties and the 90\% C.L. upper limit on m$^2_{\nu}$ for each ensemble of modified toy data sets.
The last row shows the 90\% C.L. upper limit based on toy data sets for which all systematic uncertainties were varied simultaneously.

Figure~\ref{fig:CCQE_av_logL} shows the F-C 90\% C.L. upper critical values for combined statistical and total systematic uncertainty as red dots for m$^2_{\nu-true}$ values
of 5,6 and 7 MeV$^2/c^4$ along with the average negative log-likelihood curve for the experimental data set.

\section{RESULTS}
\label{sec:Results}

With systematic uncertainties included we find a 90\% C.L. upper limit on the effective neutrino mass squared of m$^2_{\nu}  <$ 5.6 MeV$^2/c^4$.

\begin{figure}
\includegraphics[width=0.5\textwidth]{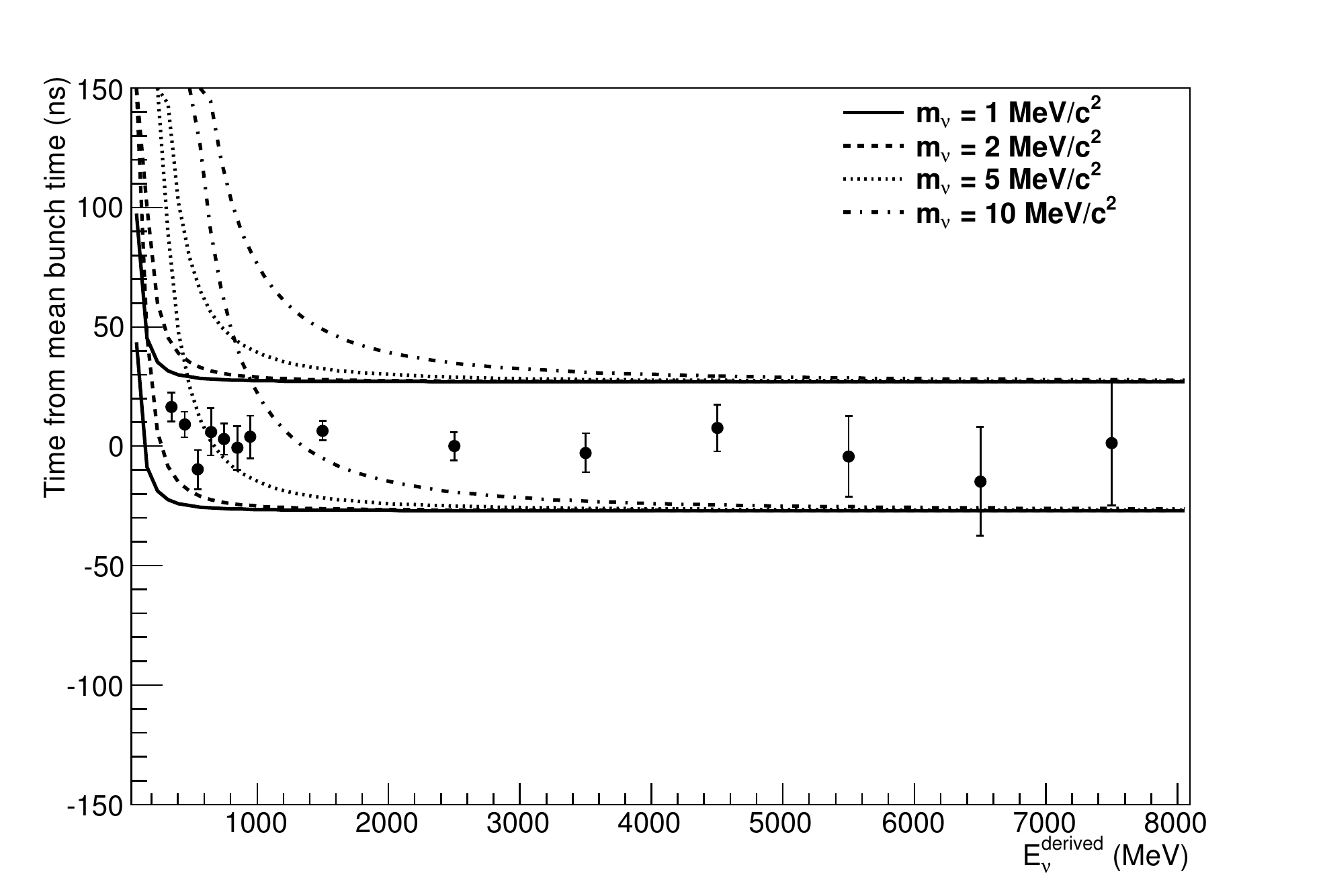} 
\caption{\label{fig:CCQE_dT_Eder_bin} Timing residuals of T2K CCQE neutrino candidate events as a function of derived neutrino energy. Events have been grouped into energy bins of 100 MeV below 1 GeV and bin sizes of 1GeV above 1 GeV. }
\end{figure}

The time deviation of individual CCQE neutrino candidate events at SK from the mean time of beam bunches as function of neutrino energy is shown 
in Fig.~\ref{fig:CCQE_dT_Eder_bin} for the 148 CCQE neutrino candidate events collected in T2K run periods I through IV. The plot was generated by assigning each event the most likely true energy, E$_{true}$ based on the MC energy conversion matrices discussed above and the events'  reconstructed energy. 
The lines indicate the expected ranges for time residuals for a variety of different effective neutrino masses. The upturn at low energies clearly shows the relativistic effect on neutrino RTOF and the dependence on effective neutrino mass.  The half-width of the bands $\sigma_{band}$ is determined by the width, $\sigma_{P_2}$, of the far detector PDF $P_2$ and its associated systematic uncertainties added in quadrature. 
Using values from tables~\ref{tab:SMRD_resol_runs}, \ref{tab:SK_RMS_res} and \ref{tab:syst_tab}  and weighting them according to the number of events in each run we determine the half-width to be $\sigma_{band}$ = 27 ns.

The vertical normalization of the data points and bands requires an absolute TOF measurement. For the present relative time of flight analysis the mean of the timing residuals for the high energy events has been adjusted to zero. At high energies and neutrino masses of a few MeV/c$^2$ or smaller no relativistic effects on the time residuals are 
expected.

The high energy tails of the calculated bands have also been centered on zero to match the adjustment of the data. We calculated and plotted bands for m$_{\nu}$ values in the range from 1 to 10 MeV$/c^2$ in Fig.~\ref{fig:CCQE_dT_Eder_bin}. Events have been grouped into energy bins of 100 MeV below 1 GeV and bin sizes of 1~GeV above 
1~GeV.

The overlap of the bands for different values of m$_{\nu}$ at high energies clearly indicates that for neutrinos with energies above 2 GeV there is no sensitivity to m$_{\nu}$  values below 10 MeV$/c^2$. All sensitivity to small values of m$_{\nu}$ stems from the energy dependence of event times for events with energies below 1 $-$ 2 GeV.

Significant improvements to the result would require timing calibration measurements at the near and far detectors to reduce systematic uncertainties.

As described in the introduction a variety of techniques to determine neutrino mass with different sensitivities exist.
Our result of an upper neutrino mass limit based on a neutrino TOF measurement may be compared 
to the MINOS 99\% C.L. upper limit on neutrino mass based on a neutrino TOF analysis of m$_{\nu} < 50$ MeV$/c^2$ \cite{MINOS_TOF}.

\section{SUMMARY}
\label{sec:Summary}

We report on a RTOF analysis based on the T2K run period I through IV data sets. The signal event sample at the far detector consists of 148 CCQE neutrino candidate events at SK for which arrival times and reconstructed energies have been determined.
A far detector PDF was fitted to the distribution of measured timing residuals of the 148 neutrino CCQE candidate events. The fit uses one free parameter, the effective neutrino mass m$_{\nu}$. The far detector PDF was constructed based on SMRD timing measurements of the beam bunch timing structure and FC$_{sideband}$ event time measurements at SK.  
	The analysis derives an upper limit on the effective neutrino mass. We find a 90\% C.L. upper limit on the effective neutrino mass square of m$^2_{\nu} < 5.6$ MeV$^2/c^4$.

\section{ACKNOWLEDGEMENTS} 

We thank the J-PARC staff for superb accelerator performance and the
CERN NA61 collaboration for providing valuable particle production data.
We acknowledge the support of MEXT, Japan;
NSERC, NRC and CFI, Canada;
CEA and CNRS/IN2P3, France;
DFG, Germany;
INFN, Italy;
National Science Centre (NCN), Poland;
RSF, RFBR and MES, Russia;
MINECO and ERDF funds, Spain;
SNSF and SER, Switzerland;
STFC, UK; and
DOE, USA.
We also thank CERN for the UA1/NOMAD magnet,
DESY for the HERA-B magnet mover system,
NII for SINET4,
the WestGrid and SciNet consortia in Compute Canada,
GridPP, UK.
In addition participation of individual researchers
and institutions has been further supported by funds from: ERC (FP7), EU;
JSPS, Japan;
Royal Society, UK;
DOE Early Career program, USA.

\bibliography{relTOF_bib}

\clearpage

\section[]{Appendix}
\label{appendix}
Table~\ref{tab:SMRD_resol_runs_detail} lists the resolutions 
$\sigma_{CT1}$ , $\sigma_{SMRD}$, $\sigma_{\Delta T}$ and $\sigma_{smrd+bunch}$
for each of the 8 (respectively 6) bunches and all four run periods. 

\begin{table}[htb] 
\caption{\label{tab:SMRD_resol_runs_detail}  SMRD and CT timing resolutions $\sigma_{CT1}$ , $\sigma_{SMRD}$, $\sigma_{\Delta T}$ and $\sigma_{smrd+bunch}$  
for run periods I through IV and each of the 6 or 8 bunches separately.}
\begin{ruledtabular}
\begin{tabular}{c c c c c c}
Run & bunch & $\sigma_{SMRD}$ &  $\sigma_{CT1}$ &  $\sigma_{\Delta T}$  &  $\sigma_{smrd+bunch}$ \\ 
period & no. & [ns] &  [ns] &  [ns] &  [ns]  \\ \hline 

\multirow{ 6}{*}{I}& 1 &    13.4  &   7.3  & 12.7   & 12 \\
& 2  &   13.9   &  8.3  &  13.1  &   12.2 \\
& 3  &   13.7  &   7.9  &  12.9   &  12.1 \\
& 4   &  13.8   &  7.8 &  12.8   &  12.1 \\
& 5   &  14.2  &  7.7  &  12.4  &   12.2 \\
& 6   &  14.4  &  9 &  12.8   &  12 \\ \hline

\multirow{ 8}{*}{IIa} & 1 &    14.5 &   9.5 & 12.6 &   11.8 \\
& 2   &  16.1  &  11.5 & 13.3  &  12.3 \\
& 3   &  16.2   & 11.3 & 14.1  &  12.9 \\
& 4   &  19.1   & 12.8 & 14.7  &  14.4 \\
& 5   &  16.7  &  16.6 & 13.1  &  9.4 \\
& 6   &  18.8  &  12 & 15.3   & 14.9 \\
& 7   &  20.5  &  16.5  &  13.9  &    13.1 \\
& 8   &  13.9  &  9 & 12.8   & 11.7 \\ \hline

\multirow{ 8}{*}{IIb} & 1 &    18.3  &   13.8 & 15.3  &   13.8 \\
& 2  &   21.2  &  16.2 & 15.5   & 14.6 \\
& 3   &  22   & 16.4 & 15.5   & 15.1 \\
& 4   &  19.9  &  14.3 & 15.9  &  14.9 \\
& 5   &  19.9  &  14.9  & 15.8   & 14.6 \\
& 6   &  21.6  &  16.3 &  15.9   & 15.1 \\
& 7   &  21.6  &  16.2 &  16   & 15.2 \\
& 8   &  21.1  &  15.8 &  16.1  &  15.1 \\ \hline

\multirow{ 8}{*}{III}  & 1 &    23 &   18.7 & 15.7 &   14.6 \\ 
& 2   &  23.5 &    19.5 &  14.9 &   14 \\
& 3   &  24.6   & 20.3 & 15.7   & 14.8 \\
& 4    & 24.3   & 19.8 &  15.6  &  14.9 \\
& 5 &  23.8   &  19.7 & 15.3   & 14.4 \\
& 6   &  24   & 19.6 &  15.5  &  14.7 \\
& 7    & 23.9  &  19.8 & 15.2  &  14.3 \\
& 8    & 23.8   & 19.9 &  14.8   & 14 \\ \hline

\multirow{ 8}{*}{IV} & 1&  14.8 & 11.6 &  14.6 & 12.2 \\
& 2 & 15.8 & 10.8 & 14.9 & 13.3 \\
& 3 & 14.7 & 9.4  &      15.2 & 13.4 \\
& 4 & 17.4 & 11.7 & 15.6 & 14.3 \\
& 5 & 18 &  12.7 &  15.4 & 14.1 \\
& 6 & 16.8 & 10.5 & 15.7 & 14.5 \\
& 7 & 17.3 & 12.6 & 15.9 & 14 \\
& 8 & 16.8 & 11.3 & 15.8 & 14.2 \\ 

\end{tabular}
\end{ruledtabular}
\end{table}

\end{document}